\documentclass[12pt]{article}
\usepackage{amsfonts,amssymb,latexsym}
\usepackage{epsfig,graphicx}
\usepackage{cite}
\newcommand{\mytitle}[1]{\large \sc #1 \\}
\newcommand{\avtor}[1]  {\large \it #1 \\}

\def\{{\lbrace}
\def\}{\rbrace}

\newtheorem{teo}{Theorem}
\newtheorem{sled}{Corollary}
\def\theequation{\arabic{section}.\arabic{equation}}

\makeatletter \@addtoreset{equation}{section}
\@addtoreset{teo}{section} \@addtoreset{sled}{teo} \makeatother

\begin{document}
\begin{center}

\mytitle{The evolution operator of the Hartree-type equation with a
quadratic potential}

\bigskip

\avtor{Lisok\footnote{Tomsk Polytechnic University, Tomsk,
Russia. E-mail: oathmat1981@mail2000.ru} A.L., Trifonov
\footnote{Tomsk Polytechnic University, Tomsk, Russia. E-mail:
trifonov@mph.phtd.tpu.edu.ru} A.Yu., Shapovalov \footnote{Tomsk
State University, Tomsk, Russia. E-mail: shpv@phys.tsu.ru}
A.V.}

\end{center}

\begin{abstract}
Based on the ideology of the Maslov's complex germ theory, a
method has been developed for finding an exact solution of the
Cauchy problem for a Hartree-type equation with a quadratic
potential in the class of semiclassically concentrated functions.
The nonlinear evolution operator has been obtained in explicit
form in the class of semiclassically concentrated functions.
Parametric families of symmetry operators have been found for the
Hartree-type equation. With the help of symmetry operators,
families of exact solutions of the equation have been constructed.
Exact expressions are obtained for the quasi-energies and their
respective states. The Aharonov-Anandan geometric phases are found
in explicit form for the quasi-energy states.
\end{abstract}


\section*{Introduction}

The integration of nonlinear equations of mathematical physics is
known to be a fundamental problem. For some classes of nonlinear
equations, families of particular solutions can be constructed
using symmetry analysis methods (see
\cite{OVS,Pukhnachev,Ibragim,OLVER,FUSCH,GAETA} and the references
therein). Of special interest are equations in multidimensional
spaces with variable coefficients. The methods of exact
integration of equations of this class are few in number.
Moreover, only a few particular cases of solving such equations
are known, which, nevertheless, are of considerable interest in
view of the complexity of the problem. For example, for a
Hartree-type equation of special form, the methods of construction
of particular solutions are discussed in \cite{Chetv}. It appears
that approximate (asymptotic) methods work well in studying the
classes of this type of equations.

A method of construction of semiclassical  asymptotics for a
Hartree-type equation with smooth coefficients and a cubic
nonlocal nonlinearity has been developed \cite{shapovalov:BTS1,
shapovalov:BTS2} based on the Maslov's complex germ theory
\cite{Maslov2, BelDob}. The key point of the method is the
integration of an auxiliary system of ordinary differential
equations -- a Hamilton-Ehrenfest system \cite{Bagre}. The
solutions of this system allow one to construct the associated
linear Schr\"odinger equation. The solutions of this equation, in
turn, make it possible to find, with a given accuracy, solutions
of the Hartree-type equation.

Besides the well-known applications in quantum mechanics (see,
e.g., \cite{Hartree,LandLif3}), the Hartree-type equation is a
base equation in constructing models which describe the
Bose--Einstein condensate (see, e.g., the review \cite
{shapovalov:PITAEVSKII}), where this equation is called the
Gross--Pitaevskii equation.

In the present work, the Cauchy problem for the one-dimensional
Hartree-type equation with a quadratic potential is solved in the
class of the semiclassically concentrated functions:
\begin{eqnarray}
&&\displaystyle\bigg\{ -i\hbar\partial_t +\widehat{\mathcal H}(t)+
\varkappa\widehat V(t,\Psi)\bigg\}\Psi =0,\label{lst03a1}\\
&&\quad\displaystyle\widehat{\mathcal H}(t)= \frac{\hat p^2}{2m} +
\frac{kx^2}{2}-eEx\cos\omega t,\cr &&\quad\displaystyle \widehat
V(t,\Psi)\Psi=\frac 12 \int\limits_{-\infty}^{+\infty}dy\,
\Big[ax^2+2bxy+cy^2\Big] |\Psi(y,t)|^2\Psi(x,t).\nonumber
\end{eqnarray}
Here,  $k>0$, $m$, $e$, $E$, $a$, $b$, and  $c$  are the
parameters of the potential, $\varkappa$ is the nonlinearity
parameter. This equation is of independent value for applications
since, for example, an external field specified by a quadratic
potential is used to describe magnetic traps in models of the
Bose-Einstein condensate. With this simple example, it is possible
to illustrate in detail the basic ideas of the method proposed in
\cite{shapovalov:BTS1, shapovalov:BTS2}. It should be stressed
that for the case under consideration this method gives an {\em
exact} solution of the Cauchy problem for the Hartree-type
equation in the class of semiclassically concentrated functions.
As a result, for a nonlinear Hartree-type equation it is possible
to find in explicit form not only the evolution operator, but also
the symmetry operators in the functional space under
consideration.

The proposed approach constructively extends the area of
application of the group analysis to the case of
quantum-mechanical integro-differential equations of the form
(\ref{lst03a1}). Symmetry operators were considered in
\cite{Pukhnachev87} for a special class of differential equations
in the framework of the $L$ -- transformation method (via a
transformation to the Lagrange variables). For these nonlinear
equations, one-parametric families of symmetry operators have also
been constructed using $L$--transformations and the invariance
group of the equation \cite{Pukhnachev87}.

It should be noted that all basic statements and constructions of
the present work remain valid, with the prescribed accuracy in $
\hbar $, $ \hbar\to 0 $, for the Hartree-type equation of general
form. A feature of the case under consideration is that all basic
statements can be checked by direct substitution, which we have
just done.

\section{The Hamilton-Ehrenfest system of equations}

The class of trajectory concentrated functions is key point in the
asymptotic integration method proposed in
\cite{shapovalov:BTS1,shapovalov:BTS2} for the Hartree-type
equation. Since the latter is a quantum mechanical equation,  the
problem of correct mathematical extraction of classical equations
of motion from quantum ones, posed by Ehrenfest \cite{Ehrenfest},
is to be studied in view of the general principles of quantum
mechanics. In respect to the correspondence principle, the
classical equations themselves enter the complete quantum
description of a system. One of the common approaches used to
attack this problem n the standard linear quantum mechanics (see,
e.g., \cite{LandLif3}) is to deduce a Hamilton--Jacobi equation
from the Schr\"odinger equation taking the formal limit
$\hbar\to0$ ($\hbar$ is the Planck constant). To obtain  solutions
of the classical Hamilton equations from the Hamilton--Jacobi
equation one has to introduce the notion of the phase trajectory
of a classical system. A mathematical framework for this approach
was developed in \cite{MasFed}. For the Hartree-type equation, it
proves to be problematic to obtain, in the above sense, the
Hamilton--Jacobi equation for a classical action. Another way to
obtain "classical" \,equations is to directly derive them from the
Hartree-type equation. To this end, a phase trajectory is to be
introduced in quantum mechanics. Let the state vector of a system
be $\Psi$ and $\hat{ x}$ and $\hat{p}$ describe a set of
generalized coordinate operators and conjugated momenta. Then
\begin{equation}
[{\hat x}, {\hat p}] = i\hbar.\label{d1.1}
\end{equation}
The quantum averages with respect to $\Psi$
\begin{equation}
\langle\hat{x}\rangle=\langle\Psi|{\hat x}|\Psi\rangle, \qquad
\langle\hat{p} \rangle=\langle\Psi|{\hat
p}|\Psi\rangle\label{d1.2}
\end{equation}
are functions of time and depend parametrically on $\hbar$
\begin{equation}
\langle\hat{x}\rangle= x_\Psi(t,\hbar), \qquad
\langle\hat{p}\rangle=p_\Psi(t,\hbar).\label{d1.3}
\end{equation}
Here, $\langle\Psi|\Phi\rangle$ is a scalar product in $L_2(\Bbb
R)$. If there exists the limit
\begin{equation}
\lim_{\hbar \to 0}x_\Psi(t,\hbar) =X(t), \qquad \lim_{\hbar \to 0}
p_\Psi(t,\hbar) = P(t),\label{d1.4}
\end{equation}
then the quantities $x=X(t)$, $p=P(t)$ are naturally to be called
the phase trajectory  of the classical system corresponding to the
state $\Psi$. Clearly, both average values (\ref{d1.3}) and the
limit values (\ref{d1.4}) depend on the state $\Psi$.
Consequently, the condition that (\ref{d1.4}) is a solution of the
classical equations of motion is a constraint on the state $\Psi$.
It is natural to speculate the solutions of the Hartree-type
equation that admit this limit to be close to classical solutions
(below we are concerned only with these solutions). Otherwise the
solutions to the Hartree-type equation are to be considered as
essentially quantum. Obviously, ${|\Psi (x,t,\hbar)|}^2$ is to
tend to $\delta\left(x-X(t)\right)$ as $\hbar \to 0$ for the
states of the first class. A similar conclusion can be made for
the wave function $\Psi$ in the $p$-representation
$\widetilde\Psi(p,t,\hbar)$, that is,
\begin{eqnarray}
&&\displaystyle\lim_{\hbar\to0}|\Psi(x,t,\hbar)|^2=\delta(x-X(t)),
\label{lst1.11a}\\
&& \displaystyle\lim_{\hbar\to0}|\widetilde\Psi( p,t,\hbar)|^2=
\delta( p-P(t)).\label{lst1.11b}
\end{eqnarray}
Let us require for a solution of the Eq. (\ref{lst03a1}) that, in
addition to conditions (\ref{lst1.11a}) and (\ref{lst1.11b}), to
moments of every order to exist. Then it is natural to seek a
solution of Eq. (\ref{lst03a1}) in the form of the following
ansatz:
\begin{equation}
\Psi(x,t,\hbar)= \varphi\Bigl(\frac{\Delta x}{\sqrt{\hbar}},t,
\sqrt{\hbar}\Bigr)
\exp\Bigl[{\frac{i}{\hbar}\Bigl(S(t,\hbar)+P(t) \Delta x
\Bigr)}\Bigr]. \label{lst03a4}
\end{equation}
Here, the function $ \varphi (\xi, t, \sqrt {\hbar}) \in
{\mathbb S} $ ($ {\mathbb S} $ is Schwartz's space) with
respect to the variable $ \xi =\Delta x/\sqrt\hbar $ and
regularly depends on $ \sqrt {\hbar} $, and $ \Delta x=x-X (t)
$. The real functions $S (t, \hbar) $, $Z (t) = (P (t), X (t))
$, which characterize the solution, are to be determined. We
shall designate the class of functions (\ref{lst03a4}) by the
symbol $ {\mathcal P} _ \hbar^t $ and call it {\em the class of
semiclassically concentrated functions} (see
\cite{shapovalov:BTS1}).

Consider the Cauchy problem
\begin{equation}
\label{lst03a2} \Psi(x,t,\hbar)\Big|_{t=s} = \psi(x,\hbar),
\quad \psi(x,\hbar)\in{\mathcal P}_\hbar^0
\end{equation}
for Eq. (\ref{lst03a1}).

For a linear operator $\widehat A$, the average value
$\langle\widehat A\rangle$ in the state $\Psi(x,t,\hbar)$ is
defined as
\begin{equation}
\label{lst03a1a} \langle  \widehat A \rangle= \frac{1}{\|\Psi
(t) \|^2}\langle\Psi(t)| \widehat A| \Psi(t)
\rangle=A_\Psi(t,\hbar).
\end{equation}
For average values of the operator $\widehat A$ on the solutions
$\Psi(t)$ of Eq. (\ref{lst03a1}) we have
\begin{eqnarray}
\label{lst03a1b} &\displaystyle \frac{d \langle \widehat
A(t)\rangle}{d t}=\Big\langle \frac{\partial \widehat
A(t)}{\partial t}
\Big\rangle+\displaystyle\frac{i}{\hbar}\langle
[\widehat{\mathcal H}_\varkappa(t,\Psi(t)),\widehat A(t)]
\rangle,
\end{eqnarray}
where $[\widehat A,\widehat B]=\widehat A\widehat B-\widehat
B\widehat A$ is the commutator of the linear operators $\widehat
A$, $\widehat B$.

By analogy with the quantum mechanical linear Schr\"odinger
equation, we shall call relation (\ref{lst03a1b}) the Ehrenfest
equation \cite{Ehrenfest}. From the Ehrenfest equation, in
particular, for $\widehat A=1$, it follows that the norm of the
solution of equation (\ref{lst03a1}) does not depend on time, that
is, we have $\|\Psi (t)\|^2 = \|\Psi (0)\|^2 = \|\Psi\|^2$. As a
result, we can pass in equation (\ref{lst03a1}), without loss of
generality, from the constant $\varkappa$ to the constant
$\tilde\varkappa =\varkappa \|\Psi \|^2 $.

Let us designate by
\begin{equation}
\alpha_\Psi^{(l,j)}(t,\hbar)=\frac{1}{\|\Psi\|^2}
\int\limits_{-\infty}^{+\infty}\Psi^*(x,t)\{(\Delta \hat p_x)^{l}
(\Delta x)^{j}\}\Psi(x,t)dx, \quad
j,l=\overline{0,\infty}\label{lst03a2a}
\end{equation}
the $j+l$-order moments of the function $\Psi(x,t)$ centered about
$x_\Psi(t,\hbar)$, $p_\Psi(t,\hbar)$. Here, $\Delta \hat p_x$
$=-i\hbar\partial_x-p_\Psi(t,\hbar)$, $\Delta
x=x-x_\Psi(t,\hbar)$, and $\{(\Delta \hat p_x)^{l}(\Delta
x)^{j}\}$ is a Weyl-ordered operator with the symbol $(\Delta
p_x)^{l}(\Delta x)^{j}$. Along with designations (\ref{lst03a2a}),
we shall use for the variances of coordinates and moments and for
the correlation functions of coordinates and moments the following
designations:
\begin{eqnarray*}
&&
\displaystyle{\sigma}_{xx}(t,\hbar)=\alpha_\Psi^{(0,2)}(t,\hbar),\quad
{\sigma}_{pp}(t,\hbar)=\alpha_\Psi^{(2,0)}(t,\hbar),\quad
{\sigma}_{xp}(t,\hbar)=\alpha_\Psi^{(1,1)}(t,\hbar).
\end{eqnarray*}

Let us write down the Ehrenfest equations for the average values
of the operators $\hat p$, $x$, $\{(\Delta x)^k(\Delta \hat
p)^l\}$, $k,l=\overline {1,M}$. As a result, we obtain for the
first-order and second-order moments the following system of
equations:
\begin{eqnarray}
&&\!\!\!\!\!\!\!\!\!\!\!\!\!\!\!\!\!\!\!\!\!\!\!\!\!\!\!\!\!\!\!\!\!
\!\!\!\!\!\!\!\!\!\left\{\begin{array}{l}
\dot{p}=-m\big(\omega_0^2+\zeta(a+b)\omega_{\rm nl}^2(a+b)\big)x+eE\cos \omega t,\\[6 pt]
\displaystyle\dot{x} =\frac{p}{m},\\[6 pt]
\end{array}\right.\label{lst03a3c1}\\ &&\!\!\!\!\!\!\!\!\!\!\!\!\!\!\!\!\!\!\!\!\!\!\!\!\!\!\!\!\!\!\!\!\!
\!\!\!\!\!\!\!\!\!\left\{\begin{array}{l}
\displaystyle\dot{\sigma}_{xx}=\frac 2m\sigma_{xp}, \\[6 pt]
\displaystyle\dot{\sigma}_{xp} = \frac
1m\sigma_{pp}-m\big(\omega_0^2+\zeta(a)\omega_{\rm nl}^2(a)
\big)\sigma_{xx},\\[6 pt]
\displaystyle\dot{\sigma}_{pp}
=-2m\big(\omega_0^2+\zeta(a)\omega_{\rm nl}^2(a)
\big)\sigma_{xp}. \\[6 pt]
\end{array}\right.\label{lst03a3c2}
\end{eqnarray}
Similarly, for the higher-order moments, we find
\begin{equation}
\left\{\begin{array}{l} \displaystyle\dot\alpha^{(j,l)}=
\frac{l}m \alpha^{(j+1,l-1)}
-jm\big(\omega_0^2+\zeta(a)\omega_{\rm nl}^2(a)\big)
\alpha^{(j-1,l+1)}+\frac{lp}m \alpha^{(j,l-1)}- \\[6 pt]
\quad\displaystyle +j\big[eE\cos \omega t-
mx\big(\omega_0^2+\zeta(a+b)\omega_{\rm nl}^2(a+b)\big)\big]
\alpha^{(j-1,l)}.
\end{array}\right.\label{lst03a3ac}
\end{equation}
Here $j,l,M\in{\mathbb N}$, $j+l=\overline{3,M}$, and we have
designated $\omega_0=\sqrt{{k}/{m}}$,\,\,\,$ \omega_{\rm nl} (u) =
\sqrt{{|\tilde\varkappa u|}/{m}}$,\,\,\,$ \zeta(u)={\rm
sign}\,\big(\tilde\varkappa u\big)$.

We shall call the system of equations (\ref{lst03a3c1}) -
(\ref{lst03a3ac}) {\em the Hamilton--Ehrenfest} system of order
$M$ ($M$ is the order of the greatest moment taken into account)
that is associated with the Hartree-type equation (\ref{lst03a1}).

Let us consider the Hamilton-Ehrenfest system (\ref{lst03a3c1}) -
(\ref{lst03a3ac}) as an abstract system of ordinary differential
equations with arbitrary initial conditions. It is obvious that
not all solutions of the Hamilton-Ehrenfest system
(\ref{lst03a3c1}) - (\ref{lst03a3ac}) can be obtained by averaging
the corresponding operators over the solutions of the Hartree-type
equation (\ref{lst03a1}). For example, the average values should
satisfy the Schr\"odinger indeterminacy relation
\begin{equation}
\sigma_{pp}\sigma_{xx}-\sigma_{xp}^2\ge
\frac{\hbar^2}4\label{lst03a4a}
\end{equation}
for the second-order moments (indeterminacy relations for
higher-order moments are given in \cite{Robertson,DoMa2}), while
the Hamilton--Ehrenfest system admits the trivial solutions $p=0$,
$x=0$, $\alpha^{(j,l)}=0$, $j+l=\overline{2,M}$. The quantity on
the left sode of relation (\ref{lst03a4a}) is the integral of
motion for the Hamilton-Ehrenfest system (\ref{lst03a3c2}) (see
\cite{bk1}). Hence, it suffices that the indeterminacy relation be
fulfilled for the time zero.

The indeterminacy relations will be fulfilled automatically if we
choose for the Hamilton-Ehrenfest system (\ref{lst03a3c1}) -
(\ref{lst03a3c2}) the following initial conditions:
\begin{equation}\begin{array}{l}
p|_{t=s}=p_0= p_\psi(\hbar), \qquad x|_{t=s}=x_0= x_\psi(\hbar)
,\\
{\sigma}_{pp}|_{t=s}=\alpha_{\psi}^{(2,0)}(\hbar),
\quad{\sigma}_{xp}|_{t=s}=\alpha_{\psi}^{(1,1)}(\hbar),
\quad{\sigma}_{xx}|_{t=s}=\alpha_{\psi}^{(0,2)}(\hbar),\\
\alpha^{(j,l)}|_{t=s}=\alpha^{(j,l)}_{\psi}, \qquad
j,l,M\in{\mathbb N},\quad j+l=\overline{3,M},
\end{array}\label{lst03a3a}
\end{equation}
where $\psi(x,\hbar)$ is the initial condition (\ref{lst03a2})
for equation (\ref{lst03a1}).

Let us designate
\begin{eqnarray}
&&\tilde\Omega=\sqrt{|\omega_0^2+\zeta(a+b)\omega_{\rm
nl}^2(a+b)|}, \label{lst03a4aa}
\quad\Omega=\sqrt{|\omega^2_0+\zeta(a)\omega_{\rm nl}^2(a)|}.
\label{lst03a4b}
\end{eqnarray}
The Hamilton--Ehrenfest system of equations
(\ref{lst03a3c1})--(\ref{lst03a3ac}) breaks into  $M$ recurrent
systems: a system of equations for the first non-centered
(initial) moments  $p_\Psi(t)$, $x_\Psi(t)$ and a system of
equations for centered moments $\alpha^{(j,l)}$ of order $n$,
$n=j+l$, $n=\overline{2,M}$. The system of order $n$ does not
depend on the moments of order above $n$.

The general solutions of systems (\ref{lst03a3c1}) and
(\ref{lst03a3c2}), depending on the sign on the expression under
the module sign in (\ref{lst03a4aa}) and (\ref{lst03a4b}), are
different in structure.
 Here we shall restrict ourselves to the case where the inequalities
 $\tilde\Omega^2 =\omega_0^2 +\zeta (a+b) \omega_{\rm nl}^2 (a+b)> 0$
 and $ \Omega^2 =\omega_0^2 +\zeta (a) \omega _ {\rm nl} ^2 (a)> 0 $
 are satisfied simultaneously.
Thus, the general solution of systems (\ref{lst03a3c1}) and
(\ref{lst03a3c2}) has the form
\begin{eqnarray}
&&\displaystyle X(t)=C_1\sin\tilde\Omega t+ C_2\cos\tilde\Omega t
+ \frac{eE}{m(\tilde\Omega^2-\omega^2)}\cos\omega t,
\nonumber \\[6 pt]&&\displaystyle
\displaystyle P(t)=m\tilde\Omega C_1\cos\tilde\Omega t
-m\tilde\Omega C_2 \sin\tilde\Omega t-
\frac{eE\omega}{(\tilde\Omega^2-\omega^2)}\sin\omega t,\label{lst03a15a}\\[6 pt]
&&\sigma_{xx}(t)=C_3\sin2\Omega t+C_4\cos2\Omega t+C_5,
\quad\sigma_{xp}(t)=m\Omega C_3\cos2\Omega
t -m\Omega C_4\sin2\Omega t,\nonumber\\[6 pt]
&&\sigma_{pp}(t)=-m^2\Omega^2 C_3\sin2\Omega t- m^2\Omega^2C_4
\cos2\Omega t+ m^2\Omega^2C_5.\label{lst03a15d}
\end{eqnarray}
Here, $C_l$, $l=\overline{1,5}$ are arbitrary constants.

Let us designate by  ${\mathfrak g}={\mathfrak g}(t,{\mathfrak
C})\in{\mathbb R}^5$ a trajectory in an extended phase space. Here
\begin{eqnarray}
\!\!\!\!\!{\mathfrak g}(t,{\mathfrak C})=\big(P(t,{\mathfrak
C}),X(t,{\mathfrak C}),\sigma_{pp}(t,{\mathfrak C}),
\sigma_{px}(t,{\mathfrak C}),\sigma_{xx}(t,{\mathfrak
C})\big)^\intercal, \quad{\mathfrak
C}=\big(C_1,C_2,C_3,C_4,C_5\big)^\intercal\label{lst03a30}
\end{eqnarray}
is the general solution of the Hamilton-Ehrenfest system
(\ref{lst03a3c1}), (\ref{lst03a3c2}) and  $\hat{\mathfrak g}$ is a
column of operators:
\begin{equation}
\hat{\mathfrak g}=\Big(\hat p, \hat x,(\Delta\hat p)^2, \frac
12(\Delta\hat p\Delta x -\Delta x\Delta\hat p),(\Delta
x)^2\big)^\intercal.\label{lst03a30a}
\end{equation}
Here,  $B^\intercal$ is the transpose of the matrix $B$. The
systems of equations (\ref{lst03a3c1})--(\ref{lst03a3c2}) can be
written as
\begin{eqnarray}
&&\dot{\mathfrak g}={\mathfrak A} {\mathfrak g}+{\mathfrak a}(t),
\qquad {\mathfrak g}\big|_{t=s}={\mathfrak g}_0, \label{lst03a30b}\\
&& {\mathfrak a}(t)=\big(eE\cos \omega t,0,0,0,0\big)^\intercal,
\nonumber
\end{eqnarray}
where
\[{\mathfrak A}=\left(\begin{array}{ccccc}
0&-m\tilde\Omega^2&0&0&0\\
\displaystyle\frac{1}{m}&0&0&0&0\\
0&0&0&-2m\Omega^2&0\\
0&0&\displaystyle\frac 1m&0&-m\Omega^2\\
0&0&0&\displaystyle\frac 2m&0
\end{array}\right),\]

\begin{teo}\label{lst_t1}
Let $\Psi(x,t)$ be a particular solution of the Hartree-type
equation {\rm (\ref{lst03a1})} with the initial condition
$\Psi(x,t)\big|_{t=0}= \psi(x)$. Let us determine the constants
${\mathfrak C}(\Psi(t))$ from the condition
\begin{equation}
{\mathfrak g}(t,{\mathfrak C})=\langle\Psi(t)|\hat{\mathfrak
g}|\Psi(t)\rangle\label{lst03a31}
\end{equation}
and the constants ${\mathfrak C}(\psi)$ from the condition
\begin{equation}
{\mathfrak g}(0,{\mathfrak C})=\langle\psi|\hat{\mathfrak g}|
\psi\rangle.\label{lst03a31a}
\end{equation}
Then we have ${\mathfrak C}(\Psi(t))={\mathfrak C}(\psi)$.
\end{teo}
{\bf Proof}. By construction, the vector
\begin{equation}
{\mathfrak g}(t)=\langle\Psi (t)|\hat{\mathfrak g}|\Psi (t)
\rangle = {\mathfrak g} (t,{\mathfrak C}(\Psi (t)))
\label{lst03a31b}
\end{equation}
is a particular solution of the system of equations
(\ref{lst03a3c1}) - (\ref{lst03a3c2}) and at the time $t=0 $ it
coincides with ${\mathfrak g}(t,{\mathfrak C}(\psi))$. By
virtue of the uniqueness of the solution of the Cauchy problem
for system (\ref{lst03a3c1}), (\ref{lst03a3c2}), the relation
\begin{equation}
{\mathfrak g}(t,{\mathfrak C}(\psi))={\mathfrak g}(t,
{\mathfrak C}(\Psi (t))), \label{lst03a31bb}
\end{equation}
is fulfilled, and thus the theorem is proved.

From equations  (\ref{lst03a3a}) and (\ref{lst03a31a}) it follows
that
\begin{equation}
\!\!{\mathfrak
C}(\psi)\!=\!\bigg(\frac{p_0}{m\tilde\Omega},x_0-\frac{e
E}{m(\tilde\Omega^2-\omega^2)},
\frac{\alpha_\psi^{(1,1)}(\hbar)}{m\Omega},
\frac{1}{2}\Big(\alpha_\psi^{(0,2)}(\hbar)-
\frac{\alpha_\psi^{(2,0)}(\hbar)}{m^2\Omega^2} \Big),
\frac{1}{2}\Big(\alpha_\psi^{(0,2)}(\hbar)+
\frac{\alpha_\psi^{(2,0)}(\hbar)}{m^2\Omega^2}\Big)\!
\bigg)^\intercal\!\!.\label{lst03a31c}
\end{equation}

\section{Associated Schr\"odinger equation}

Let us take Taylor series in $\Delta x=x-x_\Psi(t,\hbar)$,
$\Delta y=y-x_\Psi(t,\hbar)$, and  $\Delta \hat p=\hat
p-p_\Psi(t,\hbar)$ for the operators entering into equation
(\ref{lst03a1}). Then equation (\ref{lst03a1}) takes the form
\begin{eqnarray}
&&\displaystyle\{ -i\hbar\partial_t+{\mathfrak H}(t,\Psi(t))+
\langle{\mathfrak H}_z(t,\Psi(t)),\Delta\hat z\rangle
+\frac12\langle\Delta\hat z,{\mathfrak H}_{zz}(t,\Psi(t))
\Delta\hat z\rangle\}\Psi=0, \label{lst03a6}\\
&&\displaystyle{\mathfrak
H}(t,\Psi(t))=\frac{p_\Psi^2(t,\hbar)}{2m}+
\frac{kx_\Psi^2(t,\hbar)}{2}- eEx_\Psi(t,\hbar)\cos \omega t +\cr
&&\quad \displaystyle+\frac{\tilde\varkappa}2
c\alpha_\Psi^{(0,2)}(t,\hbar)+
\frac{\tilde\varkappa}2(a+2b+c)x_\Psi^2(t,\hbar),\nonumber\\
&&{\mathfrak H}_z(t,\Psi(t))= \left( \begin{array}{c}
\displaystyle\frac1m{p_\Psi(t,\hbar)}\\
 m\tilde\Omega^2x_\Psi(t,\hbar)-eE\cos \omega t\end{array}\right),
\qquad\Delta \hat z=\left(
\begin{array}{c}
\Delta \hat p \\
\Delta x\end{array}\right),\nonumber \\[6 pt] &&{\mathfrak
H}_{zz}(t,\Psi(t))=\left(
\begin{array}{cc}
\displaystyle\frac{1}{m}&0 \\
0 & m\Omega^2
\end{array}\right),\qquad z=
\left(
\begin{array}{c}
p \\[6 pt]
x\end{array}\right). \nonumber
\end{eqnarray}
Here $\langle.,.\rangle$  is denote an Euclidean scalar product of
vectors. Let us associate the nonlinear equation (\ref{lst03a6})
with the linear equation that is obtained from (\ref{lst03a6}) by
substituting the corresponding solutions of the Hamilton-Ehrenfest
system (\ref{lst03a3c1}), (\ref{lst03a3c2}) for the average values
of the operators of coordinates, momenta, and second-order
centered moments. As a  result, we obtain the following equation:
\begin{eqnarray}
&&\!\!\!\!\!\!\!\!\!\!\displaystyle\{ -i\hbar\partial_t+
{\mathfrak H}(t,\hbar,{\mathfrak g}(t,{\mathfrak C}))+
\frac12\langle{\mathfrak H}_z(t,{\mathfrak g}(t,{\mathfrak
C})), \Delta\hat z\rangle+ \langle\Delta\hat z,{\mathfrak
H}_{zz} (t,{\mathfrak g}(t,{\mathfrak C}))\Delta\hat
z\rangle\}\Phi=0,
\label{lst03a6a}\\
&&\displaystyle{\mathfrak H}(t,\hbar,{\mathfrak g}(t,{\mathfrak
C}))= \frac{P^2(t,{\mathfrak C})}{2m}+\frac{kX^2(t,{\mathfrak
C})}{2}- eEX(t,\mathfrak C)\cos \omega t +\cr &&\quad
\displaystyle+ \frac{\tilde\varkappa}2c\sigma_{xx}(t,{\mathfrak
C})+
\frac{\tilde\varkappa}2(a+2b+c)X^2(t,{\mathfrak C}),\nonumber\\
&&{\mathfrak H}_z(t,{\mathfrak g}(t,{\mathfrak C}))= \left(
\begin{array}{c}
\displaystyle\frac{1}{m}P(t,{\mathfrak C})\\
\displaystyle m\tilde\Omega^2X(t,{\mathfrak C})-eE\cos \omega t\end{array}\right),
\quad{\mathfrak H}_{zz}(t,{\mathfrak g}(t,{\mathfrak C})))=\left(
\begin{array}{cc}
\displaystyle\frac{1}{m}&0 \\
0 & m\Omega^2
\end{array}\right).\nonumber
\end{eqnarray}
We shall call equation (\ref{lst03a6a}) the {\em associated linear
Schr\"odinger equation}.

Equation(\ref {lst03a6a}) is a Schr\"odinger equation with a
quadratic Hamiltonian. It is well known that this type of equation
(see, e.g, \cite{Manko79,Perel}) admits solutions in the form of
Gaussian wave packets. Let us construct Fock's basis of the
solutions of equation (\ref{lst03a6a}) in the form accepted in the
complex germ theory \cite{Maslov2,BelDob}. By direct check we
verify that the function
\begin{eqnarray}
\Phi_0^{(0)}(x,t,{\mathfrak g}(t,{\mathfrak C}))\!=\!
N_\hbar\biggl(\displaystyle\frac{1}{C(t)}\biggr)^{1/2}\!\!\!\!\exp{\biggl\{\displaystyle\frac{i}{\hbar}
\Big(S(t,\hbar,{\mathfrak g}(t,{\mathfrak C}))+ P(t,{\mathfrak
C}) \Delta x+\frac 12\frac{B(t)} {C(t)}\Delta x^2\Big)\biggr\}}
\label{lst03a13a}
\end{eqnarray}
is a solution of equation (\ref{lst03a6a}), where
\begin{eqnarray}
&&\!\!\!\!\!\!S(t,\hbar,{\mathfrak g}(t,{\mathfrak C}))=
\displaystyle\int\limits_0^t\big(P(t,{\mathfrak C}) \dot
X(t,{\mathfrak C})-{\mathfrak H}(t,\hbar,{\mathfrak g}
(t,{\mathfrak C}))\big) dt. \label{lst03a13c}
\end{eqnarray}
Here, $B(t)$ and $C(t)$ designate, respectively, the "momentum" \,
the and "coordinate" \, parts of the solution of the system of
equations in variations
\begin{equation}
\dot a=J{\mathfrak H}_{zz}(t,{\mathfrak g}(t,{\mathfrak C}))a,
\label{lst03a13}\quad a(t)=\big( B(t),
C(t)\big)^\intercal,\label{lst03a13b}
\end{equation}
corresponding to equation (\ref{lst03a6a}).  Let us set up the
Floquet problem for the system of equations (\ref{lst03a13})
\begin{equation}
a(t+T)=e^{i\Omega T}a(t).\label{lst03a12}
\end{equation}

The solution of the Floquet problem (\ref{lst03a13}),
(\ref{lst03a12}) normalized by the condition
\begin{equation}
\{a(t), a^*(t)\}=2i, \qquad \{a_1,a_2\}=\langle a_1,J^\intercal
a_2\rangle, \label{lst03a13aa}
\end{equation}
where $J$  is the  unit symplectic matrix, has the form
\begin{equation}
a(t)=\frac{e^{i\Omega t}}{\sqrt{m\Omega}}\big( im\Omega ,
1\big)^\intercal.\label{lst03a13bb}
\end{equation}
From the normalization condition $\|\Phi_0^{(0)}(x,t,
{\mathfrak g}(t,{\mathfrak C}))\|^2=1$  we obtain
$N_\hbar=(1/\pi\hbar)^{1/4}$.

Let us designate
\begin{equation}
\hat a(t)=N_a\Big(C(t)\Delta\hat p-B(t)\Delta
x\Big).\label{lst03a43}
\end{equation}
If  $C(t)$ and  $B(t)$ are solutions of the system of equations
(\ref{lst03a13}), then the operator $\hat a(t)$ commutes with
the operator of the associated equation (\ref{lst03a6a}). Thus,
the functions
\[
\Phi_n^{(0)}(x,t,{\mathfrak g}(t,{\mathfrak
C}))=\frac{1}{\sqrt{n!}}\Big(\hat
a^+(t)\Big)^n\Phi_0^{(0)}(x,t,{\mathfrak g}(t,{\mathfrak
C}))\qquad n=\overline{0,\infty}
\]
are also solutions of the  Schr\"odinger  equation
(\ref{lst03a6a}). Commuting the operators $\hat a^+(t)$ with
the operator of multiplication by the function
$\Phi_0^{(0)}(x,t,{\mathfrak g}(t,{\mathfrak C}))$, we obtain
the following presentation for the Fock's basis of the
solutions of the associated linear equation (\ref{lst03a6a})
\begin{eqnarray*}
&\Phi_n^{(0)}(x,t,{\mathfrak g}(t,{\mathfrak
C}))=\displaystyle\frac{1}{\sqrt{n!}}
N^n_a\Phi_0^{(0)}(x,t,{\mathfrak g}(t,{\mathfrak
C}))(-i)^n[C^*(t)]^n \Big[\hbar\frac\partial{\partial x}-
\frac{2m}{|C(t)|^2}\Delta x\Big]^{n} 1=\\
&=\displaystyle\frac{1}{\sqrt{n!}}N^n_a\Phi_0^{(0)}(x,t,{\mathfrak
g}(t,{\mathfrak C}))i^n[C^*(t)]^n \biggl(\frac{\sqrt{\hbar
m}}{|C(t)|}\biggr)^n H_n \biggl(\Delta
x\frac{\sqrt{m}}{|C(t)|\sqrt\hbar}\biggr),
\end{eqnarray*}
where $H_n(\xi)$ are Hermite's polynomials. Finding
$N_a=1/\sqrt{2\hbar}$ from the condition
$$[\hat a(t),\hat a^+(t)]=1$$
and presenting the coordinate part of the solution of the system
of equations in variations in the form
\[ C(t)=\frac{1}{\sqrt{m\Omega}}\exp\big\{i\Omega t\big\}, \]
we obtain
\begin{equation}
\Phi_n^{(0)}(x,t,{\mathfrak g}(t,{\mathfrak
C}))=\frac{i^n}{\sqrt{n!}}\exp\big\{-in\Omega t\big\}
\biggl(\frac{1}{\sqrt{2}}\biggr)^n
H_n\biggl(\sqrt{\frac{m\Omega} {\hbar}}\Delta x\biggr)
\Phi_0^{(0)}(x,t,{\mathfrak g}(t,{\mathfrak
C})).\label{lst03a19}
\end{equation}
Using properties of the Hermite polynomials, we have
\begin{equation}
\alpha^{(0,2)}_{\Phi^{(0)}_n}(t,\hbar)=
\sigma_{xx}(t,\hbar,{\mathfrak C})=\frac{1}{2^nn!\sqrt{\pi}}
\int\limits_{-\infty}^\infty\Delta
x^2|\Phi_n^{(0)}(x,t,{\mathfrak g}(t,{\mathfrak C}))|^2dx=
\frac{\hbar(2n+1)}{2m\Omega}. \label{lst03a16c}
\end{equation}
We have already mentioned that the functions $\{\Phi_n^{(0)}(x,
t,{\mathfrak g}(t,{\mathfrak C}))\}_{n=0}^\infty $, while being
solutions of the associated Schr\"odinger equation
(\ref{lst03a6a}), in the general case (with arbitrary
parameters ${\mathfrak C}$) are not solutions of the Hartree
type equation (\ref{lst03a1}). The functions (\ref{lst03a19})
with fixed $n$ will be solutions of equation (\ref{lst03a1})
only if the parameters ${\mathfrak C}$ are specially chosen,
namely by the rule (\ref{lst03a31c}).

Let us put $\psi(x)=\Phi_n^{(0)}(x,0)$ in (\ref{lst03a31c}). In
view of (\ref{lst03a16c}), we obtain:
\begin{equation}
{\mathfrak C}_n={\mathfrak C}(\Phi_n^{(0)}(0))=
\bigg(\frac{p_0}{m\tilde\Omega},x_0,0,0,
\frac{\hbar(2n+1)}{2m\Omega}\bigg)^\intercal.\label{lst03a50}
\end{equation}

\begin{teo}\label{lst03at21}
For every fixed $n, n=\overline{0, \infty}$, the functions
\begin{eqnarray}
\!\!\!\!\!\!\!\!\!\!\!\!\!\!\!\Psi_n^{(0)}(x,t,{\mathfrak
g}(t,{\mathfrak C}_n))= \displaystyle\frac{i^n}{\sqrt{n!}}
e^{-in\Omega t} \biggl( \frac{1}{\sqrt{2}}\biggr)^n
H_n\biggl(\sqrt{\displaystyle \frac{m\Omega}{\hbar}}\Delta
x\biggr) \Psi_0^{(0)}
(x,t,{\mathfrak g}(t,{\mathfrak C}_n)),\label{lst03a16aa}\\
\!\!\!\Psi_0^{(0)}(x,t,{\mathfrak g}(t,{\mathfrak C}_n))\!=\!
\sqrt[4]{\displaystyle\!\frac{m\Omega}{\pi\hbar}}e^{-i\Omega
t/2}
\exp\biggl\{\displaystyle\frac{i}{\hbar}\Big(\!
S(t,\hbar,{\mathfrak g}(t,{\mathfrak C}_n))\!+\! P(t,{\mathfrak
C}_n)\Delta x\!\Big)\!-\! \frac{m\Omega}{2\hbar}\Delta
x^2\biggr\},
\end{eqnarray}
are exact solutions of equation {\rm (\ref{lst03a1})}. Here,
$\Delta x=x- X(t,{\mathfrak C}_n)$ and
\begin{eqnarray}
&&\!\!\!\!\!\!\!\!S(t,\hbar,{\mathfrak g}(t,{\mathfrak C}_n))=
\displaystyle\int\limits_0^t\big(P(t,{\mathfrak C}_n)\dot
X(t,{\mathfrak C}_n)-{\mathfrak H}(t,\hbar,{\mathfrak g}
(t,{\mathfrak C}_n))\big)dt. \label{lst03a17a}
\end{eqnarray}
The functions {\rm (\ref{lst03a16aa})} satisfy  the initial
conditions
\begin{equation}\Psi_n^{(0)}(x,t,{\mathfrak g}(t,{\mathfrak C}_n))
\Big|_{t=0}=\Phi_n^{(0)}(x,0). \label{lst03a17ag}
\end{equation}

Let us call the functions {\rm (\ref{lst03a16aa})} {\it
semiclassical trajectory coherent states} for the Hartree type
equation {\rm (\ref{lst03a1})}.
\end{teo}
{\bf Proof} of this statement is given below for a more general
case.

Associate the solutions of  the Hamilton--Ehrenfest system with an
additional periodicity condition of  with period $T=2\pi/\omega $,
to obtain
\begin{eqnarray}
\displaystyle{\mathfrak C^T_n}(\psi)\!=\!\bigg(0, \frac{e
E}{m(\tilde\Omega^2-\omega^2)}, 0, 0, \frac{\hbar(2n+1)}{2m\Omega}
\bigg)^\intercal \label{lst03a31cc}
\end{eqnarray}
and
\begin{eqnarray}
&&\!\!\!\!\!\!S(t,\hbar,{\mathfrak g}(t,{\mathfrak C}^T_n))=
\displaystyle\int\limits_0^t\big[P(t,{\mathfrak C}^T_n) \dot
X(t,{\mathfrak C}^T_n)-{\mathfrak H}(t,\hbar,{\mathfrak g}
(t,{\mathfrak C}^T_n))\big]dt=\cr
&&=\displaystyle\frac{e^2E^2}{2m(\tilde\Omega^2-\omega^2)}\Big(1+
\frac{\omega^2-\omega^2_0-\zeta(a+2b+c)\omega_{\rm nl}^2(a+2b+c)}
{2(\tilde\Omega^2-\omega^2)}\Big)t-\displaystyle
\frac{\hbar\tilde\varkappa c(2n+1)}{4m\Omega}t+\cr
&&+\displaystyle\frac{e^2E^2}{4m\omega(\tilde\Omega^2-\omega^2)}\Big(1+
\frac{-\omega^2-\omega^2_0-\zeta(a+2b+c)\omega_{\rm nl}^2(a+2b+c)}
{2(\tilde\Omega^2-\omega^2)}\Big)\sin(2\omega t).
\end{eqnarray}

In this case, the functions (\ref {lst03a16aa}) satisfy the
condition
\begin{equation}
\Psi_n^{(0)}(x,t+T,{\mathfrak C}^T_n)=e^{-i{\mathcal E}_n T/\hbar}
\Psi_n^{(0)}(x,t,{\mathfrak C}^T_n),\label{gau25b}
\end{equation}
where ${\mathcal E}$ is the quasi-energy. As a result, we obtain
for quasi-energy levels and the Aharonov-Anandan phase,
accordingly, receively
\begin{eqnarray}
&&\!\!\!\!\!\!\!\!\!\!\!\!\!\!\!{\mathcal E}_n=\displaystyle
-\frac{e^2E^2}{2m(\tilde\Omega^2-\omega^2)}
-\frac{e^2E^2[\omega^2-\omega^2_0-\zeta(a+2b+c)\omega_{\rm
nl}^2(a+2b+c)]} {4m(\tilde\Omega^2-\omega^2)^2}
+\cr&&+\hbar\Big(\Omega+ \frac{\tilde\varkappa c}{2m\Omega}\Big)
\Big(n+\frac12\Big),\label{gau27a}\\
&&\!\!\!\!\!\!\!\!\!\!\!\!\!\!\!\gamma_{\mathcal E_n} =
\displaystyle\frac{1}{\hbar}
\frac{Te^2E^2\omega^2}{2m(\tilde\Omega^2-\omega^2)^2}.
\label{gau26a}
\end{eqnarray}

\section{Nonlinear evolution operator}

Theorem \ref{lst03at21} gives a solution of the Cauchy problem
(\ref{lst03a1}), (\ref{lst03a2}) for a special class of initial
conditions of the form (\ref{lst03a17ag}). To find the Cauchy
problem solution with arbitrary initial conditions in the class of
semiclassically concentrated functions, we construct the evolution
operator for the Hartree type equation (\ref{lst03a1}).

The system of functions $\{\Phi_n^{(0)}(x,t,{\mathfrak
g}(t,{\mathfrak C}))\}^ \infty_{n=0}$ of the form (\ref{lst03a19})
constitutes a complete set of solutions of the associated linear
Schr\"odinger equation (\ref{lst03a6a}) and makes it possible to
construct the evolution operator of the Hartree type equation
(\ref{lst03a1}). The Green's function of the linear equation (the
core of the evolution operator) can be expanded over the complete
set of solutions of the linear equation:
\begin{equation}G^{(0)}(x,y,t,s,{\mathfrak g}(t,{\mathfrak C}),
{\mathfrak g}(s,{\mathfrak C}))=\sum_{n=0}^\infty\Phi_n^{(0)}
(x,t,{\mathfrak g}(t,{\mathfrak C}))
\big(\Phi_n^{(0)}(y,s,{\mathfrak g}(s,{\mathfrak C}))\big)^*
.\label{lst03a27}
\end{equation}
In view of the Meller formula \cite{Beitman2}
\[ \sum_{n=0}^\infty\frac 1{n!}\Bigl(
\frac\lambda2\Bigr)^nH_n(x)H_n(y)=
\frac1{\sqrt{1-\lambda^2}}\exp\Bigl[\frac{2xy\lambda-(x^2+y^2)\lambda^2}
{1-\lambda^2}\Bigr],\] from (\ref{lst03a19}), (\ref{lst03a27}) we
find
\begin{eqnarray*}
& \!\!\!\!\!\!G(x,y,t,s,{\mathfrak g}(t,{\mathfrak
C}),{\mathfrak g} (s,{\mathfrak
C}))\!\!=\!\sqrt{\displaystyle\frac {m\Omega}{2\pi i\hbar
\sin[\Omega(t-s)]}}\exp\!\Bigl\{\displaystyle\!\frac{i}\hbar\Bigl[
S(t,\hbar,{\mathfrak g}(t,{\mathfrak C}))\!-\!
S(s,\hbar,{\mathfrak g}(s,{\mathfrak C}))\! +
\\&+P(t,{\mathfrak C})\Delta x-P(s,{\mathfrak C})\Delta y\Bigr]
\Bigr\}\exp\Bigl\{-\displaystyle\frac {im\Omega}{2\hbar}\Big(
\frac{2\Delta x\Delta y- (\Delta x^2+\Delta
y^2)\cos[\Omega(t-s)]} {\sin[\Omega(t-s)]}\Big)\Bigr\}.
\end{eqnarray*}

\begin{teo}\label{lst03at2}
Let the operator $\widehat U_\varkappa(t,s,\cdot)$ be defined
by the relation
\begin{equation}
\widehat
U_\varkappa\big(t,s,\psi\big)(x)=\int\limits^\infty_{-\infty}
G_\varkappa\big(x,y,t,s,{\mathfrak g}(t,{\mathfrak C}(\psi)),
{\mathfrak g}(s,{\mathfrak
C}(\psi))\psi(y)\,dy,\label{lst03a28}
\end{equation}
where
\begin{eqnarray}
&&\!\!\!\!\!\!\!G_\varkappa\big(x,y,t,s,{\mathfrak g}(t,{\mathfrak
C}(\psi)), {\mathfrak g}(s,{\mathfrak C}(\psi))\big)=
\sqrt{\displaystyle\frac{m\Omega}{2\pi i\hbar
\sin[\Omega(t-s)]}}\times\cr &&\times
\exp\Bigl\{\displaystyle\frac{i}\hbar\Bigl[
S\big(t,\hbar,{\mathfrak g}(t,{\mathfrak
C}(\psi)\big)+P(t,\!{\mathfrak C}(\psi))\Delta x-
S\big(s,\hbar,{\mathfrak g}(s,{\mathfrak C}(\psi))\big)-
(s,{\mathfrak C}(\psi))\Delta y\Bigr]\Bigr\}\times\cr &&\times
\exp\Bigl\{-\displaystyle\frac
{im\Omega}{2\hbar}\Big(\frac{2\Delta x\Delta y - (\Delta
x^2+\Delta y^2)\cos[\Omega(t-s)]}
{\sin[\Omega(t-s)]}\Big)\Bigr\}\label{lst03a29}.
\end{eqnarray}
Here, $\Delta x=x-X(t,{\mathfrak C}(\psi))$, $\Delta y=
y-X(s,{\mathfrak C}(\psi))$, the function $S\big(t,\hbar,
{\mathfrak g}(t,{\mathfrak C}(\psi))\big)$ is defined by {\rm
(\ref{lst03a13c})}, and the parameters ${\mathfrak C}(\psi)$ are
determined from the equation
\begin{equation}
{\mathfrak g}(t,{\mathfrak C})\Big|_{t=s}= {\mathfrak
g}_0(\psi)=\langle\psi|\hat{\mathfrak g}|
\psi\rangle.\label{lst03a32}
\end{equation}
Then the function
\begin{equation}
\Psi (x,t)=\widehat
U_\varkappa\big(t,s,\psi\big)(x)\label{lst03a28a}
\end{equation}
is an exact solution of the Cauchy problem for the Hartree type
equation {\rm (\ref{lst03a1})} with the initial condition
$\Psi(x,t)\big|_{t=s}= \psi(x)$, and the operator $\widehat
U_\varkappa(t,s,\cdot)$ is the evolution operator for the
nonlinear Hartree type equation {\rm (\ref{lst03a1})}.
\end{teo}
{\bf Proof } can be performed by immediate substitution. Details
of calculations are given in Appendix A.

\begin{teo}\label{lst03at3}
Let the operator $\widehat U_\varkappa^{-1}(t,s,\cdot)$ be
defined by the relation
\begin{eqnarray}
&&\widehat U_\varkappa^{-1}\big(t,s,\psi\big)(x)=
\displaystyle\int\limits^\infty_{-\infty}
G_\varkappa^{-1}\big(x,y,t,s,{\mathfrak g}(t,(\psi)),
{\mathfrak g}(s,{\mathfrak C}(\psi))\big)\psi(y)\,dy=\cr
&&\qquad =\displaystyle\int\limits^\infty_{-\infty}
G_\varkappa\big(x,y,s,t,{\mathfrak g}(s,{\mathfrak C}(\psi)),
{\mathfrak g}(t,(\psi))\big)\psi(y) \,dy.\label{lst03a33}
\end{eqnarray}
Here, the function $G_\varkappa\big(x,y,s,t,{\mathfrak
g}(s,{\mathfrak C}(\psi)), {\mathfrak g}(t,(\psi))\big)$ is
defined by relation {\rm (\ref{lst03a28})} in which the
variable $t$ should be replaced by $s$ and the variable $s$ by
$t$; the constants ${\mathfrak C}$ are determined from the
equation
\begin{equation}
{\mathfrak g}(s,{\mathfrak C})\Big|_{s=t}={\mathfrak
g}_0(\psi)= \langle\psi|\widehat{\mathfrak
g}|\psi\rangle.\label{lst03a32a}
\end{equation}
The operator $\widehat U_\varkappa^{-1}(t,s,\cdot)$ {\rm
(\ref{lst03a33})} is then the left inverse one to the operator
$\widehat U_\varkappa(t,s,\cdot)$ {\rm (\ref{lst03a28})};
that is,
\begin{eqnarray}
&&\widehat U_\varkappa^{-1}\big(t,s,\widehat
U_\varkappa\big(t,s,\psi\big)\big)(x)=\psi(x),\qquad
\psi\in{\mathcal P}_\hbar^0.\label{lst03a37}
\end{eqnarray}
\end{teo}
{\bf Proof } can be performed by immediate substitution. Details
of calculations are given in  Appendix B.
\begin{sled}\rm
If the function  $\Psi(x,t)$ is a particular solution of the
Hartree type equation (\ref{lst03a1}), then
\begin{eqnarray}
&&\widehat U_\varkappa\big(t,s,\widehat
U_\varkappa^{-1}\big(t,s,\Psi(t)\big)\big)(x)=\Psi(x,t).
\label{lst03a37a}
\end{eqnarray}
\end{sled}
{\bf Proof.} Let us designate:  $\psi(x)=\Psi(x,t)\big|_{t=s}$.
Then by virtue of theorem  \ref{lst03at2}, we have
$\Psi(x,t)=\widehat U_\varkappa\big(t,s,\psi\big)(x)$. Therefore,
the left side of relation (\ref{lst03a37a}) can be presented in
the form
\begin{eqnarray*}
\widehat U_\varkappa\big(t,s,\widehat
U_\varkappa^{-1}\big(t,s,\Psi(t)\big)\big)(x)=
\widehat U_\varkappa\big[t,s,\hat U_\varkappa^{-1}\big(t,s,
\widehat U_\varkappa\big(t,s,\psi\big)\big)\big](x)=
\widehat U_\varkappa\big(t,s,\psi\big)(x)=\Psi(x,t).
\end{eqnarray*}
Here, we used formula (\ref{lst03a37}). So, the statement is
proved.

\section{Symmetry operators for the Hartree type equation}

The solution of the Cauchy problem for the Hartree type equation
{\rm (\ref{lst03a1})} in the class of semiclassically concentrated
functions ${\mathcal P}^0_\hbar$ and the explicit form of the
evolution operator (\ref{lst03a28}) allow one, in turn, to
construct in explicit form the general expressions for the basic
constructions of symmetry analysis
\cite{OVS,Pukhnachev,Ibragim,FUSCH,GAETA,OLVER}. These
constructions are: symmetry operators, a one-parametric family of
symmetry operators, and generators of this family (symmetries of
equation (\ref{lst03a1})).

Actually, let $\hat{\textsf {a}}$ be some operator acting in
${\mathcal P}^0_\hbar$, ($ \hat {\textsf{a}}: {\mathcal P}
^0_\hbar\to{\mathcal P}^0_\hbar $) and $\Psi (x,t)$ is any
function from the class ${\mathcal P}^t_\hbar$. Let us define
the operator $\widehat{\textsf {A}}(\cdot)$ by the relation
\begin{eqnarray}
&&\Phi(x,t)=\widehat{\textsf{A}}\big(\Psi(t)\big)(x)= \widehat
U_\varkappa\big(t,\hat {\textsf{a}}\, \widehat
U_\varkappa^{-1}\big(t,\Psi(t)\big)\big)(x), \label{lst03a38}
\end{eqnarray}
where $ \widehat U_\varkappa(t,\cdot)=\widehat
U_\varkappa(t,0,\cdot)$.

\begin{teo}\label{lst03at6}
If the function $\Psi(x,t)$ is a solution of the  Hartree type
equation {\rm (\ref{lst03a1})}, then  $\Phi(x,t)$ {\rm
(\ref{lst03a38})} is also the solution of equation {\rm
(\ref{lst03a1})}.
\end{teo}
{\bf Proof } immediately follows from Theorem \ref{lst03at2} and
Theorem \ref{lst03at3}.

\bigskip
Thus, the operator {\rm $\widehat{\textsf{A}}(\cdot)$}, determined
by relation {\rm (\ref{lst03a38})} is the symmetry operator for
equation (\ref{lst03a1}).

Now let the operator $\hat {\textsf{b}}$ and its operator exponent
$\exp(\alpha\hat {\mathfrak b})$ act in the class ${\mathcal
P}^0_\hbar$; that is, $\hat {\textsf{b}}: {\mathcal P}^0_\hbar
\to{\mathcal P}^0_\hbar$ and $\exp(\alpha\hat
{\textsf{b}}):{\mathcal P}^0_\hbar \to{\mathcal P}^0_\hbar$, where
$\alpha$ is a real parameter. For an arbitrary function
$\Psi(x,t)\in{\mathcal P}^t_\hbar$, let us define a one-parametric
family of operators $\widehat{\textsf{B}}(\alpha,\cdot)$ by the
relation
\begin{eqnarray}
&&\widehat{\textsf{B}}\big(\alpha,\Psi(t)\big)(x)= \widehat
U_\varkappa\big(t,\exp\{\alpha\hat{\textsf{b}}\} \widehat
U_\varkappa^{-1}\big(t,\Psi(t)\big)\big)(x). \label{lst03a39}
\end{eqnarray}

By analogy with the above constructions, the operators
$\widehat{\textsf{B}}(\alpha,\cdot)$ constitute the one-parametric
family of symmetry operators of equation (\ref{lst03a1}).

It is easy to verify that for an arbitrary function
$\Psi(x,t)\in{\mathcal P}^t_\hbar$  the group property is valid:
\begin{eqnarray}
&&\widehat{\textsf{B}}\big(\alpha+\beta,\Psi(t)\big)(x)=
\widehat{\textsf{B}}\Big(\alpha,\widehat
{\textsf{B}}\big(\beta,\Psi(t)\big)\Big)(x). \label{lst03a40}
\end{eqnarray}

Differentiating relation (\ref {lst03a39}) with respect to the
parameter $\alpha$ at the point $\alpha=0$, we obtain
\begin{equation}
\widehat{\textsf{C}}\big(\Psi(t)\big)(x)=\frac{d}{d \alpha}
\widehat{\textsf{B}}\big(\alpha,\Psi(t)\big)(x)\Big|_{\alpha=0}=
\frac{d}{d\alpha}\widehat U_\varkappa\big(t,\exp\{
\alpha\hat{\textsf{b}}\} \widehat U_\varkappa^{-1}
\big(t,\Psi(t)\big)\big)(x)\Big|_{\alpha=0}.\label{lst03a41}
\end{equation}
The operator $\widehat{\textsf {C}}(\cdot)$, defined by
relation (\ref{lst03a41}), is the generator of the
one-parameter  family of symmetry operators (\ref {lst03a40}).

Note that the operator $\widehat{\textsf{C}}(\cdot)$ is not a
symmetry operator of equation (\ref{lst03a1}). This is due to the
fact that the parameters ${\mathfrak C}$ that determine the
evolution operator $ \widehat U_\varkappa(t,\cdot)$ in relation
(\ref{lst03a41}) depend on $\alpha $. Actually, the quantities
${\mathfrak C}$ are determined from equation (\ref{lst03a32}):
\begin{eqnarray*}
&{\mathfrak g}(t,{\mathfrak
C})\Big|_{t=0}=\big\langle\exp\{\alpha
\hat{\textsf{b}}\}\phi\big| \hat{\mathfrak g}\big|
\exp\{\alpha\hat {\textsf{b}}\}\phi\big\rangle,\quad
\phi(x)=\widehat U_\varkappa^{-1}\big(t,\Psi(t)\big)(x),\quad
\Psi(x,t)\in{\mathcal P}^t_\hbar,
\end{eqnarray*}
which contains the parameter $\alpha$ in explicit form. Thus,
expression (\ref{lst03a41}) will contain the derivative of the
evolution operator $\widehat U_\varkappa (t,\cdot)$ with
respect to the parameters ${\mathfrak C}$. This derivative is
an operator other than the evolution operator.

{\bf Example}. Let us substitute into the relation
(\ref{lst03a38}), instead of the operators  $\hat {\textsf{a}}$,
the operators $\hat a^+(t)$ and $\hat a(t)$ of the form
(\ref{lst03a43}) for $t=0$. Here,
\begin{eqnarray*}
&\displaystyle\hat a(0)=\frac1{\sqrt{2\hbar m\Omega}}\big[
\Delta\hat p_0-im\Omega\Delta x_0\big], \qquad \hat
a^+(0)=\frac1{\sqrt{2\hbar m\Omega}}\big[ \Delta\hat
p_0+im\Omega\Delta x_0\big],
\end{eqnarray*}
where $\Delta\hat p_0=-i\hbar\partial_x-p_0$ and $\Delta
x_0=x-x_0$. Then the operators $\widehat A^{(\pm)}(\cdot)$
determined by the relations
\begin{eqnarray*}
\widehat A^{(+)}\big(\Psi(t)\big)(x)= \widehat
U_\varkappa\big(t,\hat a^+(0)\,
\widehat U_\varkappa^{-1}\big(t,\Psi(t)\big)\big)(x),
~~\widehat A^{(-)}\big(\Psi(t)\big)(x)= \widehat
U_\varkappa\big(t,\hat a(0)\, \widehat
U_\varkappa^{-1}\big(t,\Psi(t)\big)\big)(x),
\end{eqnarray*}
where $\Psi(x,t)\in{\mathcal P}^t_\hbar$, are the symmetry
operators of equation (\ref{lst03a1}). For these operators, we
have in particular
\begin{eqnarray*}
&&\widehat A^{(+)}\big(\Psi_n^{(0)}(t,{\mathfrak g}
(t,{\mathfrak C}_n))\big)(x)
=\sqrt{n+1}\,\Psi_{n+1}^{(0)}(x,t,{\mathfrak g}(t,{\mathfrak C}_{n+1})),\\
&&\widehat A^{(-)}\big(\Psi_n^{(0)}(t,{\mathfrak g}
(t,{\mathfrak C}_n))\big)(x)
=\sqrt{n}\,\Psi_{n-1}^{(0)}(x,t,{\mathfrak g}(t,{\mathfrak
C}_{n-1})).
\end{eqnarray*}
Here, $\Psi_n^{(0)}(x,t,{\mathfrak g}(t,{\mathfrak C}_n))$ are
semiclassical trajectory coherent states of the form
(\ref{lst03a16aa}), where constants ${\mathfrak C}_n$ are
defined by  relation (\ref{lst03a50}). Therefore, the operators
 $\widehat A^{(+)}(\cdot)$ are nonlinear analogs
 of "creation -- annihilation" operators.
With the help of the operators $\widehat A^{(\pm)}(\cdot)$,
relations (\ref{lst03a16aa}) and  (\ref{lst03a50}) can be
presented in the form
\begin{equation}
\Psi_n^{(0)}(x,t,{\mathfrak g}(t,{\mathfrak C}_n))=
\frac{1}{\sqrt{n!}} \bigl(\widehat A^{(+)}(\cdot)\bigr)^n
\Psi_0^{(0)}(x,t,{\mathfrak g}(t,{\mathfrak
C}_0)).\label{lst03a47}
\end{equation}

Let us define the one-parametric family of shift operators
$\widehat D(\alpha,\cdot)$ for the functions
$\Psi(x,t)\in{\mathcal P}^t_\hbar$ by the relation
\begin{equation}
\widehat D\big(\alpha,\Psi(t)\big)(x)=\widehat U_\varkappa
\big(t,\hat{\mathcal D}_0(\alpha)\,\widehat U_\varkappa^{-1}
\big(t,\Psi(t)\big)\big)(x),\label{lst03a45}
\end{equation}
where
\begin{eqnarray}
& \widehat{\mathcal D}_0(\alpha)=\exp\{ \alpha\hat{a}{}^+(0)-
\alpha^*\hat{a}(0))\},\label{lst03a45a}\qquad
\Psi(x,t)\in{\mathcal P}^t_\hbar,\qquad \alpha\in{\mathbb C}.
\end{eqnarray}

The operators $\widehat D(\alpha,\cdot)$ are the symmetry
operators for  equation (\ref{lst03a1}), and the functions
\begin{equation}
\Psi_\alpha(x,t,{\mathfrak g}(t,{\mathfrak C}_\alpha))= \widehat
D\big(\alpha,\Psi_0^{(0)}(t,{\mathfrak g} (t,{\mathfrak
C}_0))\big)(x)\label{lst03a46}
\end{equation}
are the solutions of equation (\ref{lst03a1}) for arbitrary
complex values of $\alpha$. Here, the parameters ${\mathfrak
C}_\alpha$ are defined by  equation (\ref{lst03a32}), which, for
this case has the form
\begin{equation}
{\mathfrak g}(t,{\mathfrak C})\Big|_{t=0}=
\big\langle\widehat{\mathcal D}_0(\alpha)\Psi_0^{(0)}(0)\big|
\hat{\mathfrak g}\big|\widehat{\mathcal D}_0(\alpha)
\Psi_0^{(0)}(0)\big\rangle.\label{lst03a51}
\end{equation}
Let us write the operator  $\widehat{\mathcal D}_0(\alpha)$ as
\begin{eqnarray}
\displaystyle\widehat{\mathcal D}_0(\alpha)=\exp
\{\beta\Delta\hat{p}_0+ \gamma\Delta x_0\}=\displaystyle\exp
\Big\{-\displaystyle\frac{i\hbar}2 \beta\gamma\Big\}
\exp\{\gamma\Delta
x_0\}\exp\{\beta\Delta\hat{p_0}\},\label{lst03prf10}
\end{eqnarray}
where $ \beta=\Big[\alpha- \alpha^*\Big]/{\sqrt{2\hbar m\Omega}}$,
$\gamma =i\Big[ \alpha+\alpha^*\Big]\sqrt{{m\Omega}/{2\hbar}}$.
Therefore,
\begin{eqnarray}
&\!\!\!\!\!\!\Psi_\alpha(x,0,{\mathfrak g}(0,{\mathfrak
C}_\alpha))= \widehat{\mathcal
D}_0(\alpha)\Psi_0^{(0)}(x,0,{\mathfrak g} (0,{\mathfrak
C}_0))= \cr & =\displaystyle\exp \Big\{-\frac{i\hbar}2
\beta\gamma\Big\} \exp\{\gamma\Delta x_0- p_0\beta\}
\Psi_0^{(0)}(x-i\hbar\beta,0, {\mathfrak g}(0,{\mathfrak
C}_\alpha)).\label{lst03prf12}
\\&\displaystyle
\!\!\!\!\!\Psi_0^{(0)}(x-\!i\hbar\beta,0,\!{\mathfrak
g}(0,{\mathfrak C}_\alpha))\!=
\!\!\displaystyle\Psi_0^{(0)}(x,0,{\mathfrak g}(0,{\mathfrak
C}_\alpha))\exp \!\Big\{\!\!\frac{i}\hbar \Big[\! -i\hbar
p_0\beta - i\hbar Q(0) \Delta x_0 \beta-\frac {\hbar^2} 2 Q(0)
\beta^2\!\Big]\!\!\Big\}.\nonumber\label{lst03prf13}
\end{eqnarray}
Notice that  $Q(0)=im\Omega$ and, consequently, $\gamma+Q(0)\beta
= 2i\alpha\sqrt{{m \Omega}/{2\hbar}}$. Similar manipulations yield
\begin{equation}
-\frac{i\hbar} 2 [\beta\gamma + Q(0)\beta^2]=\sqrt{\frac {\hbar
m\Omega} 2}\alpha \beta=\frac 12
\big[\alpha^2-|\alpha|^2\big].\label{lst03prf15}
\end{equation}
Substituting  
(\ref{lst03prf15}) into  (\ref{lst03prf12}), we obtain
\begin{eqnarray}
& \Psi_\alpha(x,0,{\mathfrak g}(0,{\mathfrak C}_\alpha))
=\Psi_0^{(0)}(x,0,{\mathfrak g}(0,{\mathfrak C}_0))\exp\Big\{
-\displaystyle\frac{|\alpha|^2}2+i\sqrt{\displaystyle\frac{2m\Omega}\hbar}
\alpha\Delta x_0+\frac {\alpha^2}2\Big\}=\cr
&=\sqrt[4]{\displaystyle\frac{m\Omega}{\pi\hbar}}\exp\Big\{
-\displaystyle\frac{|\alpha|^2}2+\frac{\alpha^2}2+
\frac{i}{\hbar}\big(p_0+ \sqrt{2m\Omega\hbar}\alpha\big)\Delta
x_0 -\frac{m\Omega}{2\hbar}\Delta x_0^2
\Big\}.\label{lst03prf16}
\end{eqnarray}
Denoting  $\alpha_1={\rm Re}\,\alpha$ and $\alpha_2={\rm
Im}\,\alpha$, we can write
\begin{equation}
|\Psi_\alpha(x,0,{\mathfrak g}(0,{\mathfrak C}_\alpha))|^2
=\sqrt{\frac{m\Omega}{\pi\hbar}}\exp\Big\{
-\frac{m\Omega}\hbar\Big(\Delta x_0
+\sqrt{\frac{2\hbar}{m\Omega}}\alpha_2\Big)^2
\Big\}.\label{lst03a54}
\end{equation}

From equation (\ref{lst03a51}), in view of (\ref{lst03a54}), we
obtain
\begin{equation}
{\mathfrak C}_\alpha=\bigg(\frac{1}{m\tilde\Omega}p_\alpha,
x_\alpha,0,0,\frac{\hbar}{2m\Omega}\bigg)^\intercal,\label{lst03a52}
\end{equation}
where $ p_\alpha=p_0+\alpha_1\sqrt{2m\Omega\hbar}$, $
x_\alpha=x_0-\alpha_2\sqrt{{2\hbar}/{m\Omega}}$.

As a result, we find an explicit form of the functions
$\Psi_\alpha(x,t,{\mathfrak g}(t,{\mathfrak C}_\alpha))$ for the
one-parametric family (with the parameter $\alpha$) of solutions
for the nonlinear Hartree type equation (\ref{lst03a1}). Thereby,
on direct application of the one-parametric family of symmetry
operators $\widehat D(\alpha,\cdot)$ to the function
$\Psi_0(x,t,{\mathfrak g}(t,{\mathfrak C}_0))$ we obtain the
explicit expressions
\begin{equation}
\Psi_\alpha(x,t,{\mathfrak g}(t,{\mathfrak C}_\alpha))=
\sqrt[4]{\frac{m\Omega}{\pi\hbar}}\exp\biggl\{
\frac{i}{\hbar}\Big(S(t,\hbar,{\mathfrak g}(t,{\mathfrak
C}_\alpha))+ P(t,{\mathfrak C}_\alpha)\Delta x\Big)-
\frac{m\Omega}{2\hbar}\Delta x^2\biggr\},\label{lst03a53}
\end{equation}
where $\Delta x=x- X(t,{\mathfrak C}_\alpha)$, and
\begin{eqnarray}
&&\!\!\!\!\!\!\!\!S(t,\hbar,{\mathfrak g}(t,{\mathfrak
C}_\alpha))= \displaystyle\int\limits_0^t\big(P(t,{\mathfrak
C}_\alpha)\dot X (t,{\mathfrak C}_\alpha)-{\mathfrak
H}(t,\hbar,{\mathfrak g} (t,{\mathfrak C}_\alpha))\big)dt.
\label{lst03a55}
\end{eqnarray}

For the operators (\ref{lst03a45}), the following multiplication
low is valid:
\begin{equation}
\widehat D\Big(\alpha,\widehat D\big(\beta,\Psi(t)\big)\Big)
(x)=\exp[\alpha\beta^*-\alpha^*\beta]\widehat D
\big(\alpha+\beta,\Psi(t)\big)(x). \label{lst03a44}
\end{equation}
The operators $\exp(i\gamma)\widehat D(\alpha,\cdot)$, where
$\gamma\in\mathbb R$, $\alpha\in\mathbb C$, determine the
nonlinear analog of the Heisenberg--Weyl group representation
\cite{Manko79,Perel}.  The function $\Psi_\alpha(x,t,{\mathfrak g}
(t,{\mathfrak C}_\alpha))$ (\ref{lst03a53}), by virtue of
(\ref{lst03a15d}) and (\ref{lst03a52}), minimize the Schr\"odinger
uncertainty relation (\ref{lst03a4a}) and, hence, thay are
compressed coherent states.
\bigskip
To summarize, we note that the exact expressions constructed in
this work for the evolution operator and symmetry analysis
constructions for the Hartree type equation (\ref{lst03a1}) can be
generalized for the case of a Hartree type equation in a
multidimensional space with smooth coefficients of general form.
This can be done based on the results of \cite{shapovalov:BTS1,
shapovalov:BTS2}. However, this generalization will be valid only
in the sense of approximation to within $\widehat
O(\hbar^{(M+1)/2})$, $\hbar\to 0$, where $M$ is the order of the
Hamilton - Ehrenfest system. In particular, this allows one to
construct a special kind of approximate symmetry operators (and
symmetries) for the above Hartree type equations, which are
natural to be called semiclassical symmetry operators
(symmetries).

\section*{Appendix A}
\def\theequation{{\rm A}.\arabic{equation}}
\setcounter{equation}{0}

{\bf Proof of the theorem \ref{lst03at2}} 1. Let us calculate the
limit of the function  $\Psi(x,t)$ as $t\to s+0$ . We have
\begin{eqnarray*}
& \displaystyle\lim_{t\to s+0}\Psi (x,t)=\lim_{t\to s+0}
\int\limits^\infty_{-\infty}G_\varkappa\big(x,y,t,s,{\mathfrak g}
(t,{\mathfrak C}(\psi)),{\mathfrak g}
(s,{\mathfrak C}(\psi))\big)\psi(y)\,dy=
\int\limits^\infty_{-\infty}dy\,\psi(y) \times\\
&\displaystyle\!\!\!\times\lim_{t\to
s+0}\sqrt{\!\displaystyle\frac{m\Omega} {2\pi i\hbar
\sin[\Omega(t-s)]}} \!\exp\Bigl\{\!\displaystyle\frac{i}\hbar
\Bigl[\!S\big(t,\hbar,{\mathfrak g}(t,{\mathfrak C}(\psi)\big)-
S\big(s,\hbar,{\mathfrak g}(s,{\mathfrak C}(\psi))\!\big)
+\!P(t,{\mathfrak C}(\psi))\Delta x-\\&-\displaystyle
P(s,{\mathfrak C}(\psi)) \Delta y\Bigr] \Bigr\}\exp\Bigl\{-\frac
{im\Omega}{2\hbar} \Big(\frac{2\Delta x\Delta y-(\Delta x^2+\Delta
y^2)
\cos[\Omega(t-s)]} {\sin[\Omega(t-s)]}\Big)\Bigr\}=\\
&=\displaystyle\int\limits^\infty_{-\infty}dy\, \psi(y)\lim_{t\to
s+0} \sqrt{\displaystyle\frac{m}{2\pi i\hbar (t-s)}} \exp\Bigl\{
\frac{im}{2\hbar}\frac{(\Delta x-\Delta y)^2} {t-s}\Bigr\}=
\int\limits^\infty_{-\infty}dy\psi(y)\delta(x-y)=\psi(x).
\end{eqnarray*}
Therefore,  $\Psi(x,t)\big|_{t=s}= \psi(x)$.

2.  Recall that for the solutions of the Hartree type equation
(\ref{lst03a1}), the relation (see (\ref{lst03a31bb}))
\begin{equation}
{\mathfrak g}(t,{\mathfrak C}(\psi))=\langle\Psi(t)|
\hat{\mathfrak g}|\Psi(t)\rangle\label{lst03a32c}
\end{equation}
is valid.

Since it is not proved yet that the functions $\Psi (x,t)$
(\ref{lst03a28a}) are the solution of equation (\ref{lst03a1}),
while and relation (\ref{lst03a32c}) is used below, we verify its
validity by direct check.

The definition of moments (\ref{lst03a2a}) involves the norm of
the function $\Psi(x,t)$ (\ref{lst03a28a}). Let us calculate
this norm as follows:
\begin{eqnarray*}
&\|\Psi(t)\|^2=\displaystyle\int\limits^\infty_{-\infty}dx
\int\limits^\infty_{-\infty}dy
\int\limits^\infty_{-\infty}dz\psi(y)\psi^*(z)
\sqrt{\displaystyle\frac{m\Omega}{2\pi
i\hbar\sin[\Omega(s-t)]}}
\sqrt{\displaystyle\frac{m\Omega}{2\pi i\hbar
\sin[\Omega(t-s)]}}
\times\\
&\times\exp\Bigl\{-\displaystyle\frac{i}\hbar\Bigl[
S\big(t,\hbar,{\mathfrak g}(t,{\mathfrak C}(\psi)\big)-
S\big(s,\hbar,{\mathfrak g}({\mathfrak C}(\psi))\big)
+P(t,{\mathfrak C}(\psi))\Delta x-P(s,{\mathfrak
C}(\psi))\Delta z
\Bigr]\Bigr\}\times\\
&\times\exp\Bigl\{-\displaystyle\frac{im\Omega}{2\hbar}
\Big(\frac{2\Delta x\Delta z-(\Delta x^2+\Delta z^2)
\cos[\Omega(t-s)]}
{\sin[\Omega(s-t)]}\Big)\Bigr\}\exp\Bigl\{\displaystyle\frac{i}\hbar\Bigl[
S\big(t,\hbar,{\mathfrak g}(t,{\mathfrak C}(\psi)\big)-\\&-
S\big(s,\hbar,{\mathfrak g}(s,{\mathfrak C}(\psi))\big)+
P(t,{\mathfrak C}(\psi))\Delta x-P(s,{\mathfrak C}(\psi))\Delta
y
\Bigr] \Bigr\}\times\\
&\times\exp\Bigl\{\displaystyle\frac {im\Omega}{2\hbar}
\Big(\frac{2\Delta x\Delta y-(\Delta x^2+\Delta y^2)
\cos[\Omega(s-t)]} {\sin[\Omega(s-t)]}\Big)\Bigr\}=\\
&=\displaystyle\int\limits^\infty_{-\infty}dx
\int\limits^\infty_{-\infty}dy \int\limits^\infty_{-\infty}dz
\frac{m\Omega}{2\pi \hbar \sin[\Omega(s-t)]}\exp\Bigl\{
\frac{i}\hbar\Bigl[P(s,{\mathfrak C}(\psi))
(\Delta z-\Delta y)\Bigr]\times\\
&\times\exp\Bigl\{\displaystyle\frac{im\Omega}{2\hbar}
\Big(\frac{2(\Delta z-\Delta y)\Delta x-(\Delta z^2-\Delta y^2)
\cos[\Omega(t-s)]}{\sin[\Omega(t-s)]}\Big)\Bigr\}\psi(y)\psi^*(z)=\\
&=\displaystyle\int\limits^\infty_{-\infty}dy
\int\limits^\infty_{-\infty}dz\delta(z-y)\exp\Bigl\{
\frac{i}\hbar\Bigl[P(s,{\mathfrak C}(\psi))(z-y)\Bigr]\times\\
&\times\exp\Bigl\{\displaystyle\frac{im\Omega}{2\hbar}
\Big(\frac{(\Delta y^2-\Delta z^2)\cos[\Omega(t-s)]}
{\sin[\Omega(t-s)]}\Big)\Bigr\}\psi(y)\psi^*(z)=
\int\limits^\infty_{-\infty}dx\,|\psi(x)|^2.
\end{eqnarray*}
Thus, we obtain $\|\Psi(t)\|=\|\psi\|$.

Let us show that
\begin{equation}
x_\Psi(t,\hbar)=X(t,{\mathfrak C}(\psi)).\label{lst03a28ac}
\end{equation}

Calculate the non-centered moment of the first order
(\ref{lst03a28ac}). Using the explicit form of the functions
$\Psi(x,t)$, we find
\begin{eqnarray*}
&&x_\Psi(t,\hbar)=\displaystyle\frac1{\|\Psi(t)\|^2}
\int\limits^\infty_{-\infty}dx\int\limits^\infty_{-\infty}dy
\int\limits^\infty_{-\infty}dz x \sqrt{\displaystyle\frac
{m\Omega}{2\pi i\hbar \sin[\Omega(t-s)]}}
\sqrt{-\displaystyle\frac{m\Omega}{2\pi i\hbar
\sin[\Omega(t-s)]}} \times\\
&&\quad\times\exp\Bigl\{-\displaystyle\frac{i}\hbar\Bigl[
S\big(t,\hbar,{\mathfrak g}(t,{\mathfrak C}(\psi)\big)-
S\big(s,\hbar,{\mathfrak g}(s,{\mathfrak C}(\psi))\big)
+P(t,{\mathfrak C}(\psi))\Delta x-
P(s,{\mathfrak C}(\psi))\Delta z\Bigr] \Bigr\}\times\\
&&\quad\times\exp\Bigl\{-\displaystyle\frac
{im\Omega}{2\hbar}\Big( \frac{2\Delta x\Delta z-(\Delta
x^2+\Delta z^2)\cos[\Omega(t-s)]}
{\sin[\Omega(s-t)]}\Big)\Bigr\}\times\\
&&\quad\times\exp\Bigl\{\displaystyle\frac{i}\hbar\Bigl[
S\big(t,\hbar,{\mathfrak g}(t,{\mathfrak C}(\psi)\big)-
S\big(s,\hbar,{\mathfrak g}(s,{\mathfrak C}(\psi))\big)+
P(t,{\mathfrak C}(\psi))\Delta x-
P(s,{\mathfrak C}(\psi))\Delta y\Bigr] \Bigr\} \times\\
&&\quad\times\exp\Bigl\{\displaystyle\frac{im\Omega}{2\hbar}
\Big(\frac{2\Delta x\Delta y-(\Delta x^2+\Delta
y^2)\cos[\Omega(s-t)]}
{\sin[\Omega(s-t)]}\Big)\Bigr\}\psi(y)\psi^*(z)=\\
&&\quad=\displaystyle\int\limits^\infty_{-\infty}dx
\int\limits^\infty_{-\infty}dy \int\limits^\infty_{-\infty}dz
\frac{m\Omega(X(t,{\mathfrak C}(\psi))+\Delta
x)\psi(y)\psi^*(z)}
{2\pi \hbar \sin[\Omega(t-s)]}\times\\
&&\quad\times\exp\Bigl\{\displaystyle\frac i\hbar\Bigl[
P(s,{\mathfrak C}(\psi))(\Delta z-\Delta y)\Bigr] \Bigr\}
\exp\Bigl\{\frac{im\Omega}{\hbar}\Big(\frac{\Delta x
(\Delta y-\Delta z)}{\sin[\Omega(s-t)]}\Big)\Bigr\}\times\\
&&\quad\times\exp\Bigl\{\displaystyle\frac{im\Omega}{2\hbar}\Big(
\frac{(\Delta z^2 -\Delta y^2)\cos[\Omega(s-t)]}
{\sin[\Omega(s-t)]}\Big)\Bigr\}=X(t,{\mathfrak C}(\psi))+\\
&&\quad +\displaystyle\int\limits^\infty_{-\infty}dz
\int\limits^\infty_{-\infty}dy
\int\limits^\infty_{-\infty}d\omega
\frac{\psi(y)\psi^*(z)}{2\pi}\Bigl( \frac{\hbar\omega
\sin[\Omega(t-s)]}{m\Omega}\Bigr)
\exp\Bigl\{i\omega(\Delta y-\Delta z)\Bigr\} \times\\
&&\quad\times\exp\Bigl\{\displaystyle\frac{i}\hbar\Bigl[
P(s,{\mathfrak C}(\psi))(\Delta z-\Delta y)\Bigr] \Bigr\}
\exp\Bigl\{\frac{im\Omega}{2\hbar} \frac{(\Delta z^2-\Delta
y^2)\cos[\Omega(s-t)]}
{\sin[\Omega(s-t)]}\Bigr\}=\\
&&\quad=X(t,{\mathfrak C}(\psi))+
\displaystyle\int\limits^\infty_{-\infty}dz
\int\limits^\infty_{-\infty}dy\,\delta'(y-z)\psi(y)\psi^*(z)
\frac{\hbar\sin[\Omega(t-s)]}{m\Omega} \times\\
&&\quad\times\exp\Bigl\{\displaystyle\frac{i}\hbar\Bigl[
P(s,{\mathfrak C}(\psi))(\Delta z-\Delta y)\Bigr]\Bigr\}
\exp\Bigl\{\frac {im\Omega}{2\hbar}\Big( \frac{(\Delta
z^2-\Delta y^2)\cos[\Omega(s-t)]}
{\sin[\Omega(s-t)]}\Big)\Bigr\}=\\
&&\quad= X(t,{\mathfrak C}(\psi))+\displaystyle\frac12
\int\limits^\infty_{-\infty}dz\Bigg\{
\frac{\hbar\sin[\Omega(t-s)]}{m\Omega} \exp\Bigl\{\frac{i}\hbar
P(s,{\mathfrak C}(\psi))
(\Delta z-\Delta y)\Bigr\} \times\\
&&\quad\times\exp\Bigl\{\displaystyle\frac {im\Omega}{2\hbar}
\frac{(\Delta z^2-\Delta y^2)\cos[\Omega(s-t)]}
{\sin[\Omega(s-t)]}\Bigr\}\psi(y)\psi^*(z)\Bigg\}'_y\Bigg|_{y=z}+\\
&&\quad+\displaystyle\frac12\int\limits^\infty_{-\infty}dy
\Bigg\{\frac{\hbar\sin[\Omega(t-s)]}{m\Omega}\exp\Bigl\{
-\frac{i}{\hbar}P(s,{\mathfrak C}(\psi))(\Delta z-\Delta y)
\Bigr\} \times\\
&&\quad\times\exp\Bigl\{\displaystyle\frac {im\Omega}{2\hbar}
\frac{(\Delta z^2-\Delta y^2)\cos[\Omega(s-t)]}
{\sin[\Omega(s-t)]}\Bigr\}\psi(y)\psi^*(z)\Bigg\}_z'
\Bigg|_{z=y}= X(t,{\mathfrak C}(\psi))+\\
&&\quad
+\displaystyle\frac12\frac{\hbar\sin[\Omega(t-s)]}{m\Omega}
\int\limits^\infty_{-\infty}dz\Big\{-\frac{i}\hbar
P(s,{\mathfrak C}(\psi))\psi(z)\psi^*(z)+\psi'(z)\psi^*(z)\Big\}+\\
&&\quad+\displaystyle\frac12\frac{\hbar\sin[\Omega(t-s)]}{m\Omega}
\int\limits^\infty_{-\infty}dy\Big\{\frac{i}\hbar
P(s,{\mathfrak C}(\psi))\psi(y)\psi^*(y)+\psi(y)(\psi^*(y))'
\Big\}=\\
&&\quad=X(t,{\mathfrak C}(\psi))+ \displaystyle\frac{\hbar
\sin[\Omega(t-s)]}{m\Omega}
\int\limits^\infty_{-\infty}dy\Big\{\psi(y)(\psi^*(y))\Big\}'=
X(t,{\mathfrak C}(\psi)).
\end{eqnarray*}
Thus, relation (\ref{lst03a28ac}) is valid.

Let us show that
\begin{equation}
\alpha_\Psi^{(0,2)}(t,\hbar)= \alpha^{(0,2)}(t,\hbar,{\mathfrak
C}(\psi)).\label{lst03a28aa}
\end{equation}
Calculate the moment of the second order (\ref{lst03a28aa}). Using
the explicit form of the functions $\Psi(x,t)$ we obtain
\begin{eqnarray*}
&\displaystyle\alpha_\Psi^{(0,2)}(t,\hbar)=\frac1{\|\Psi(t)\|^2}
\int\limits^\infty_{-\infty}dx\int\limits^\infty_{-\infty}dy
\int\limits^\infty_{-\infty}dz (\Delta x)^2
\sqrt{\displaystyle\frac{m\Omega}{2\pi i\hbar
\sin[\Omega(t-s)]}} \sqrt{\displaystyle\frac{-m\Omega}{2\pi
i\hbar\sin[\Omega(t-s)]}}
\times\\
&\displaystyle\times\exp\Bigl\{-\frac{i}\hbar\Bigl[
S\big(t,\hbar,{\mathfrak g}(t,{\mathfrak C}(\psi)\big)-
S\big(s,\hbar,{\mathfrak g}(s,{\mathfrak C}(\psi))\big)
+P(t,{\mathfrak C}(\psi))\Delta x-P(s,{\mathfrak C}(\psi))
\Delta z\Bigr]\Bigr\}\times\\
&\displaystyle\times\exp\Bigl\{-\frac{im\Omega}{2\hbar}
\Big(\frac{2\Delta x\Delta z-(\Delta x^2+\Delta z^2)
\cos[\Omega(t-s)]}{\sin[\Omega(s-t)]}\Big)\Bigr\}\times\\
&\displaystyle\times\exp\Bigl\{\frac{i}\hbar\Bigl[
S\big(t,\hbar,{\mathfrak g}(t,{\mathfrak C}(\psi)\big)-
S\big(s,\hbar,{\mathfrak g}(s,{\mathfrak C}(\psi))\big)+
P(t,{\mathfrak C}(\psi))\Delta x-P(s,{\mathfrak C}(\psi))
\Delta y\Bigr] \Bigr\} \times\\
&\displaystyle\times\exp\Bigl\{\frac {im\Omega}{2\hbar}
\Big(\frac{2\Delta x\Delta y-(\Delta x^2+\Delta
y^2)\cos[\Omega(s-t)]}
{\sin[\Omega(s-t)]}\Big)\Bigr\}\psi(y)\psi^*(z) =\\
&=\displaystyle\int\limits^\infty_{-\infty}dx
\int\limits^\infty_{-\infty}dy \int\limits^\infty_{-\infty}dz
\frac{m\Omega(\Delta x)^2\psi(y)\psi^*(z)} {2\pi \hbar
\sin[\Omega(t-s)]}\exp\Bigl\{-\frac{i}\hbar\Bigl[
P(s,{\mathfrak
C}(\psi))(\Delta z-\Delta y)\Bigr] \Bigr\}\times\\
&\displaystyle\times \exp\Bigl\{\frac{im\Omega}{\hbar}\Big(
\frac{\Delta x(\Delta y-\Delta z)}{\sin[\Omega(s-t)]}
\Big)\Bigr\}
\exp\Bigl\{\frac{im\Omega}{2\hbar}\Big( \frac{(\Delta
z^2-\Delta y^2)\cos[\Omega(s-t)]}
{\sin[\Omega(s-t)]}\Big)\Bigr\}=\\
&=-\displaystyle\int\limits^\infty_{-\infty}dz
\int\limits^\infty_{-\infty}dy
\int\limits^\infty_{-\infty}d\omega
\frac{\psi(y)\psi^*(z)}{2\pi}\Bigl(\frac{\hbar\omega
\sin[\Omega(t-s)]}{m\Omega}\Bigr)^2 \exp\Bigl\{
i\omega(\Delta y-\Delta z)\Bigr\} \times\\
&\displaystyle\times\exp\Bigl\{\frac{i}\hbar\Bigl[
P(s,{\mathfrak C}(\psi))(\Delta y-\Delta z)\Bigr] \Bigr\}
\exp\Bigl\{\frac{im\Omega}{2\hbar}\frac{(\Delta z^2-\Delta y^2)
\cos[\Omega(s-t)]} {\sin[\Omega(s-t)]}\Bigr\}=\\
&=-\displaystyle\int\limits^\infty_{-\infty}dz
\int\limits^\infty_{-\infty}dy\,\delta''(y-z)\psi(y)\psi^*(z)
\frac{\hbar^2\sin^2[\Omega(t-s)]}{m^2\Omega^2}
\exp\Bigl\{-\frac{i}\hbar\Bigl[
P(s,{\mathfrak C}(\psi))(\Delta z-\Delta y)\Bigr] \Bigr\}\times\\
&\displaystyle\times\exp\Bigl\{\frac{im\Omega}{2\hbar}
\Big(\frac{(\Delta z^2-\Delta y^2)\cos[\Omega(s-t)]}
{\sin[\Omega(s-t)]}\Big)\Bigr\}=
-\int\limits^\infty_{-\infty}dz\Bigg\{\psi(y)\psi^*(z)
\frac{\hbar^2\sin^2[\Omega(t-s)]}{m^2\Omega^2}\times\\
&\displaystyle\times\exp\Bigl\{-\frac{i}\hbar P(s,{\mathfrak
C}(\psi))(\Delta z-\Delta y)\Bigr\} \exp\Bigl\{\frac
{im\Omega}{2\hbar}\frac{(\Delta z^2-\Delta y^2)
\cos[\Omega(s-t)]}{\sin[\Omega(s-t)]}\Bigr\}
\Bigg\}_{zz}''\Bigg|_{y=z}.
\end{eqnarray*}
The expression in braces can be rearranged as follows:
\begin{eqnarray*}
&&\Bigg\{\displaystyle\frac{\hbar^2\sin^2[\Omega(t-s)]}{m^2\Omega^2}
\exp\Bigl\{-\frac{i}\hbar\Bigl[P(s,{\mathfrak C}(\psi))
(\Delta z-\Delta y)\Bigr] \Bigr\}\times\\
&&\quad\times\exp\Bigl\{\displaystyle\frac{im\Omega}{2\hbar}
\Big(\frac{(\Delta z^2-\Delta y^2)\cos[\Omega(s-t)]}
{\sin[\Omega(s-t)]}\Big)\Bigr\}\psi(y)\psi^*(z)\Bigg\}''_{y=z}=\\
&&\quad=-\displaystyle\frac{\hbar^2 \sin^2[\Omega(t-s)]}
{m^2\Omega^2}\Bigg\{-\frac{p_0^2}{\hbar^2}\psi(z)\psi^*(z)+
\frac{2p_0}{\hbar^2}m\Omega\Delta z\,{\rm ctg}\,[\Omega(t-s)]
\psi(z)\psi^*(z)-\\
&&\quad-\displaystyle\frac{2i}\hbar p_0\psi'(z)\psi^*(z)-
\frac{2i}\hbar m\Omega\,{\rm ctg}\,[\Omega(s-t)]\Delta
z\psi'(z)\psi^*(z)+\psi''(z)\psi^*(z)-\\
&&\quad-\displaystyle\frac{1}{\hbar^2} m^2\Omega^2\, {\rm
ctg}^2\Omega(s-t)]\Delta z^2\psi(z)\psi^*(z)-
\frac{im\Omega}\hbar\,{\rm ctg}\,
[\Omega(s-t)]\psi(z)\psi^*(z)\Bigg\}= \\
&&\quad=\displaystyle\frac{p_0^2}{m^2\Omega^2}\sin^2[\Omega(s-t)]
\psi(z)\psi^*(z)+ \frac{2p_0}{m\Omega}\sin[\Omega(s-t)]
\cos[\Omega(s-t)]\Delta z\psi(z)\psi^*(z)+\\
&&\quad+\displaystyle\frac{2i\hbar p_0}{m^2\Omega^2}
\sin^2[\Omega(s-t)]\psi'(z)\psi^*(z)+
\frac{2i\hbar}{m\Omega}\sin[\Omega(s-t)]
\cos[\Omega(s-t)]\Delta z\psi'(z)\psi^*(z)-\\
&&\quad-\displaystyle\frac{\hbar^2}
{m^2\Omega^2}\sin^2[\Omega(s-t)]\psi''(z)\psi^*(z)
+\cos^2\Omega(s-t)]\Delta z^2\psi(z)\psi^*(z)+\\
&&\quad+ \displaystyle\frac{i\hbar}{m\Omega}\sin[\Omega(s-t)]
\cos[\Omega(s-t)]\psi(z)\psi^*(z).
\end{eqnarray*}
Then,
\begin{eqnarray*}
&\alpha_\Psi^{(0,2)}(t,\hbar) =
\displaystyle\frac{\sin^2[\Omega(s-t)]}{m^2\Omega^2}\Big[ p_0^2
\int\limits^\infty_{-\infty}dz\,\psi^*(z)\psi(z)  -2 p_0
\displaystyle\int\limits^\infty_{-\infty}dz\,
\psi^*(z)(-i\hbar)\psi'(z)
+\\&+\displaystyle\int\limits^\infty_{-\infty}dz\,
\psi^*(z)(-i\hbar)^2\psi''(z)\Big]+\displaystyle\frac{\sin[2\Omega(s-t)]}{m\Omega}\Big[p_0
\int\limits^\infty_{-\infty}dz\,\psi^*(z)\Delta z\psi(z)-\\&-
\displaystyle\int\limits^\infty_{-\infty}dz\, \psi^*(z)\Delta
z(-i\hbar)\psi'(z) +\displaystyle\frac{i\hbar}2
\int\limits^\infty_{-\infty}dz\,
\psi^*(z)\psi(z)\Big]+\cos^2\Omega(s-t)]
\int\limits^\infty_{-\infty}dz\,\Delta z^2\psi^*(z)\psi(z).
\end{eqnarray*}
In view of the relations
\begin{eqnarray*}
&&\alpha^{(2,0)}_\psi=\displaystyle\int\limits^\infty_{-\infty}dz
\,(-i\hbar\partial_z-p_0)^2\psi^*(z)\psi(z),\qquad
\alpha^{(0,2)}_\psi=\displaystyle\int\limits^\infty_{-\infty}
dz\,\Delta z^2\psi^*(z)\psi(z),\\
&&\alpha^{(1,1)}_\psi=\displaystyle\int\limits^\infty_{-\infty}dz
\frac12\big[\Delta z(-i\hbar\partial_z-p_0)
+(-i\hbar\partial_z-p_0)\Delta z\big]\psi^*(z)\psi(z)=\\
&&\quad=- p_0 \displaystyle\int\limits^\infty_{-\infty}dz\,
\psi^*(z)\Delta z\psi(z)+\int\limits^\infty_{-\infty}dz\,
\psi^*(z)\Delta z(-i\hbar)\psi'(z) -
\frac{i\hbar}2 \int\limits^\infty_{-\infty}dz\,\psi^*(z)\psi(z),
\end{eqnarray*}
in the case  $s=0$, we obtain
\begin{eqnarray*}
&&\alpha_\Psi^{(0,2)}(t,\hbar) =-\alpha^{(2,0)}_\psi
\displaystyle\frac{\sin^2[\Omega(t)]}{m^2\Omega^2}
-\alpha^{(1,1)}_\psi\frac{\sin[2\Omega(t)]}{m\Omega}
+\alpha^{(0,2)}_\psi\cos^2\Omega(s-t)]=\\
&&\quad=\displaystyle\frac{\alpha_\psi^{(1,1)}}{m\Omega}
\sin2\Omega t+\frac{1}{2}\Big(\alpha_\psi^{(0,2)}
-\frac{\alpha_\psi^{(2,0)}}{m^2\Omega^2} \Big)\cos2\Omega t
+\frac{1}{2}\Big(\alpha_\psi^{(0,2)}+
\frac{\alpha_\psi^{(2,0)}}{m^2\Omega^2}\Big).
\end{eqnarray*}

From here, in view of formulas (\ref{lst03a15d}) and
(\ref{lst03a31c}), relation (\ref{lst03a28aa}) follows. The proof
of other equalities in (\ref{lst03a32c}) is similar.

3. Let us show that the function  $\Psi(x,t)$ satisfies the
equation  (\ref{lst03a1}). Substituting (\ref{lst03a28a}) into
(\ref{lst03a1}), we obtain
\begin{eqnarray*}
&\displaystyle\bigg\{ -i\hbar\partial_t +\widehat{\mathcal
H}(t)+ \varkappa\widehat V(t,\Psi)\bigg\}\Psi(x,t,\hbar)
=\int\limits^\infty_{-\infty}\,dy\psi(y)\bigg\{
-i\hbar\partial_t
+\frac{\hat p^2}{2m}+\frac{kx^2}{2} +\\
&+\displaystyle\frac{\tilde\varkappa} 2
\Big[ax^2+2bx[x_\Psi(t,\hbar)+\alpha_\Psi^{(0,1)}(t,\hbar)]+
c[x_\Psi^2(t,\hbar)+
2x_\Psi(t,\hbar)\alpha_\Psi^{(0,1)}(t,\hbar)
+\alpha_\Psi^{(0,2)}(t,\hbar)])\Big]\bigg\}\times\\
&\times G_\varkappa\big(x,y,t,s,{\mathfrak g}(t,{\mathfrak
C}(\psi)), {\mathfrak g}_0(\psi)\big)=
\displaystyle\int\limits^\infty_{-\infty}dy\,\psi(y)
G_\varkappa\big(x,y,t,s,{\mathfrak g}(t,{\mathfrak C}(\psi)),
{\mathfrak g}_0(\psi)\big)\times\\
&\times\bigg\{\displaystyle\frac{i\hbar\Omega}2
\frac{\cos[\Omega(t-s)]}{\sin[\Omega(t-s)]}+ \dot
S\big(t,\hbar,{\mathfrak g}(t,{\mathfrak C}(\psi))\big)+ \dot
P(t,{\mathfrak C}(\psi))\Delta x -
P(t,{\mathfrak C}(\psi))\dot X(t,{\mathfrak C}(\psi))-\\
&- {m\Omega\dot X(t,{\mathfrak C}(\psi))}
\displaystyle\frac{\Delta y-\Delta x\cos[\Omega(t-s)]}
{\sin[\Omega(t-s)]} - \frac{m\Omega^2}{2}\Big(\frac{2\Delta
x\Delta y\cos[\Omega(t-s)] -
(\Delta x^2+\Delta y^2)} {\sin^2[\Omega(t-s)]}\Big)+\\
& + \displaystyle\frac{1}{2m} \Big[ P^2(t,{\mathfrak
C}(\psi))+2P(t,{\mathfrak C}(\psi)){m\Omega} \frac{\Delta y -
\Delta x\cos[\Omega(t-s)]}
{\sin[\Omega(s-t)]}+ \\
&+\Big({m\Omega}\displaystyle\frac{\Delta y-\Delta x
\cos[\Omega(t-s)]} {\sin[\Omega(s-t)]}\Big)^2\Big]-
\frac{i\Omega\hbar}{2}\frac{\cos[\Omega(t-s)]}
{\sin[\Omega(t-s)]} +\frac{(k+\tilde\varkappa a)}{2}
\big(X^2(t,{\mathfrak C}(\psi))+\\
& \displaystyle+2X^2(t,{\mathfrak C}(\psi))\Delta x+\Delta
x^2\big) + \tilde\varkappa b\big(X(t,{\mathfrak
C}(\psi))+\Delta
x\big)[x_\Psi(t,\hbar)+\alpha_\Psi^{(0,1)}(t,\hbar)]+\\
&\displaystyle+\frac{\tilde\varkappa c} 2[x_\Psi^2(t,\hbar)+
2x_\Psi(t,\hbar)\alpha_\Psi^{(0,1)}(t,\hbar)
+\alpha_\Psi^{(0,2)}(t,\hbar)] \bigg\}.
\end{eqnarray*}
Using relations  (\ref{lst03a32c}) (where, in particular,
$\alpha_\Psi^{(0,1)}(t,\hbar)=0$) and the expressions
\begin{eqnarray*}
&&\dot S\big(t,\hbar,{\mathfrak g}(t,{\mathfrak C}(\psi))\big)=
P(t,{\mathfrak C}(\psi))\dot X(t,{\mathfrak C}(\psi))- 
\displaystyle\frac{P^2(t,{\mathfrak C}(\psi))}{2m}
+\frac{kX^2(t,{\mathfrak C}(\psi))}{2}+\\
&&\quad+\tilde\varkappa c\sigma_{xx}(t,\hbar,{\mathfrak
C}(\psi))+ \displaystyle\frac{\tilde\varkappa}2(a+2b+c)
X^2(t,{\mathfrak C}(\psi))
\end{eqnarray*}
we obtain
\begin{eqnarray*}
&&\big\{ -i\hbar\partial_t +\widehat{\mathcal H}(t)+\varkappa
\widehat V(t,\Psi)\big\}\Psi(x,t,\hbar) =\\
&&\quad=\displaystyle\int\limits^\infty_{-\infty}dy\,\psi(y)
G_\varkappa\big(x,y,t,s,{\mathfrak g}(t,{\mathfrak C}(\psi)),
{\mathfrak g}_0(\psi)\big)
\bigg\{\dot P(t,{\mathfrak C}(\psi))\Delta x -\\
&&\quad -\displaystyle\frac{m\Omega^2}2(\Delta x)^2+
\frac{(k+\tilde\varkappa a)(2X(t,{\mathfrak C}(\psi)) \Delta
x+(\Delta x)^2)}{2}+\tilde\varkappa bX^2(t,{\mathfrak C}(\psi))
\Delta x \bigg\}=0,
\end{eqnarray*}
Q. E. D.

\section*{Appendix B}
\def\theequation{{\rm B}.\arabic{equation}}
\setcounter{equation}{0}

{\bf Proof of Theorem \ref{lst03at3}.} Assume the opposite, that
is, relation (\ref{lst03a37}) is not valid and involves  some
function $\Phi (x,t,s)$ on the right side. Then, in view of
(\ref{lst03a28}), relation (\ref{lst03a37}) can be represented as
$\Phi (x,t,s) = \widehat U_\varkappa^{-1}\big (t,s,\Psi\big)(x)$,
{\rm where} \quad $\Psi (x,t)=\widehat U_\varkappa\big
(t,s,\psi\big)(x)$. According to the definition of the operator
$\widehat U_\varkappa^{-1}(t,s,\cdot)$ (\ref{lst03a33}), the
parameters \linebreak ${\mathfrak C} (\Psi (t))$ entering in this
relation are determined from equation (\ref{lst03a32a}), which in
our case becomes
\begin{equation}
{\mathfrak g}(s,{\mathfrak C})\Big|_{s=t}= \langle\Psi
(t)|\hat{\mathfrak g}| \Psi(t)\rangle.\label{lst03a32b}
\end{equation}
By virtue of Theorem \ref{lst_t1}, the relation ${\mathfrak C}(\Psi (t))
 = {\mathfrak C}(\psi)$ is valid.
 Hence, the operator $\widehat U_\varkappa^{-1}(t,s,\cdot)$,
 as it acts on the function $\Psi (x,t)$, and the operator
 $\widehat U_\varkappa^{-1}(t,s,\cdot)$, as it acts on the function
 $\psi (x)$, are defined on the same trajectory
 ${\mathfrak g}(t,{\mathfrak C}(\psi))$.
 Designate $\Delta x=x-X(s,{\mathfrak C}(\psi))$,
 $\Delta y=y-X (t, {\mathfrak C}(\psi))$, $\Delta z=z-X (s,{\mathfrak C}(\psi))$;
 then (\ref{lst03a37}) can be presented as
\begin{eqnarray*}
&\!\!\!\Phi(x,t,s)= \!\widehat
U_\varkappa^{-1}\big(t,s,\widehat
U_\varkappa\big(t,s,\psi\big)\big)(x)\!=\!\displaystyle\int\limits^\infty_{-\infty}\!\!dy\!
\int\limits^\infty_{-\infty}\!\!dz\!\sqrt{\displaystyle\frac
{m\Omega}{2\pi i\hbar \sin[\Omega(s-t)]}}
\sqrt{\displaystyle\frac{m\Omega}{2\pi i\hbar
\sin[\Omega(t-s)]}}
\times\\
&\times\exp\Bigl\{\displaystyle\frac{i}\hbar\Bigl[
S\big(s,\hbar,{\mathfrak g}(s,{\mathfrak C}(\psi))\big)-
S\big(t,\hbar,{\mathfrak g}({\mathfrak C}(\psi))\big)+
P(s,{\mathfrak C}(\psi))\Delta x-P(t,{\mathfrak C}(\psi))
\Delta z\Bigr] \Bigr\}\times\\
&\times\exp\Bigl\{\displaystyle\frac {im\Omega}{2\hbar}
\Big(\frac{2\Delta x\Delta z-(\Delta x^2+\Delta z^2)
\cos[\Omega(t-s)]}{\sin[\Omega(t-s)]}\Big)\Bigr\}\times\\
&\times\exp\Bigl\{\displaystyle\frac{i}\hbar\Bigl[
S\big(t,\hbar,{\mathfrak g}(t,{\mathfrak C}(\psi)\big)-
S\big(s,\hbar,{\mathfrak g}(s,{\mathfrak C}(\psi))\big)
+P(t,{\mathfrak C}(\psi))\Delta z-P(s,{\mathfrak C}(\psi))
\Delta y\Bigr] \Bigr\}\times\\
&\times\exp\Bigl\{\displaystyle\frac{im\Omega}{2\hbar}
\Big(\frac{2\Delta z\Delta y-(\Delta z^2+\Delta y^2)
\cos[\Omega(s-t)]}{\sin[\Omega(s-t)]}\Big)\Bigr\}\psi(y)=\\
&=\displaystyle\int\limits^\infty_{-\infty}dy
\int\limits^\infty_{-\infty}dz\frac {m\Omega}{2\pi \hbar
\sin[\Omega(s-t)]}\exp\Bigl\{\frac{i}\hbar\Bigl[
P(s,{\mathfrak C}(\psi))(\Delta x-\Delta y)\Bigr]\Bigr\}\times\\
&\times\exp\Bigl\{\displaystyle\frac{im\Omega}{2\hbar}
\Big(\frac{2(\Delta x-\Delta y)\Delta z-(\Delta x^2-\Delta y^2)
\cos[\Omega(t-s)]} {\sin[\Omega(t-s)]}\Big)\Bigr\}\psi(y)=
\displaystyle\int\limits^\infty_{-\infty}dy\,\delta(x-y)\times\\&\times
\displaystyle\exp\Bigl\{\frac i\hbar\Bigl[P(s,{\mathfrak
C}(\psi))(x-y)
\Bigr]\exp\Bigl\{\displaystyle\frac{im\Omega}{2\hbar}
\Big(\frac{(\Delta y^2-\Delta x^2)\cos[\Omega(t-s)]}
{\sin[\Omega(t-s)]}\Big)\Bigr\}\psi(y)=\psi(x).
\end{eqnarray*}
and this contradiction proves the theorem.

\bigskip
The work has been  supported in part by President of the Russian
Federation grants  NSh-1743.2003.2 and MD-246.2003.02; Ministry of
Education of the Russian Federation grant N А03-2.8-794; A.L.
Lisok has been recipient the scholarship of the non-commertial
Fond "Dinastija".

\end{document}